\documentclass{aa}

\usepackage[varg]{txfonts}
\usepackage{hyperref}
\usepackage[usenames]{color}
\hypersetup{dvips, colorlinks=true, linkcolor=black, citecolor=black, filecolor=blue, urlcolor=blue}
\usepackage{graphicx}
\bibpunct{(}{)}{;}{a}{}{,}

\begin{document}

\title{Shock-accelerated cosmic rays and streaming instability in the adaptive mesh refinement code Ramses}

\titlerunning{CR acceleration and streaming instability}
\authorrunning{Y. Dubois et al.}

\author{Yohan Dubois\inst{1}, Beno\^it Commer\c con\inst{2}, Alexandre Marcowith\inst{3} \and Loann Brahimi\inst{3}}
\institute{Institut d'Astrophysique de Paris, UMR 7095, CNRS, UPMC Univ. Paris VI, 98 bis boulevard Arago, 75014 Paris, France \\\email{dubois@iap.fr}
\and
 Centre de Recherche Astrophysique de Lyon UMR5574, ENS de Lyon, Univ. Lyon1, CNRS, Universit\'e de Lyon, 69007 Lyon, France 
 \and
 Laboratoire Univers et Particules de Montpellier (LUPM), Universit\'e Montpellier, CNRS/IN2P3, CC72, place Eug\`ene Bataillon,
34095 Montpellier Cedex 5, France }

\date{Received 10 July 2019 / Accepted 19 September 2019}

\abstract{Cosmic rays (CRs) are thought to play a dynamically important role in several key aspects of galaxy evolution, including the structure of the interstellar medium, the formation of galactic winds, and the non-thermal pressure support of halos. We introduce a numerical model solving for the CR streaming instability and acceleration of CRs at shocks with a fluid approach in the adaptive mesh refinement code {\sc ramses}. CR streaming is solved with a diffusion approach and its anisotropic nature is naturally captured. We introduce a shock finder for the {\sc ramses} code that automatically detects shock discontinuities in the flow. Shocks are the loci for CR injection, and their efficiency of CR acceleration is made dependent on the upstream magnetic obliquity according to the diffuse shock acceleration mechanism. We show that the shock finder accurately captures shock locations and estimates the shock Mach number for several problems. The obliquity-dependent injection of CRs in the Sedov solution leads to situations where the supernova bubble exhibits large polar caps (homogeneous background magnetic field), or a patchy structure of the CR distribution (inhomogeneous background magnetic field). Finally, we combine both accelerated CRs with streaming in a simple turbulent interstellar medium box, and show that the presence of CRs significantly modifies the structure of the gas.}


\maketitle

\section{Introduction}

Cosmic rays (CR) are understood to  play an important role in astrophysical plasmas due to their capacity to ionise the interstellar matter \citep{padovanietal09} and their non-negligible pressure support to gas dynamics according to evolutionary processes that differ substantially from the thermal component since they diffuse efficiently and have different dissipation timescales. CRs are likely produced at shocks through the process of diffuse shock acceleration (DSA)~(see \citealp{bell78,drury83,blandford&eichler87,jones&ellison91,berezhko&ellison99} and~\citealp{marcowithetal16} for a recent review). Recent advances in the numerical modelling of DSA through hybrid particle-in-cell codes~\citep{caprioli&spitkovski14acc} have provided accurate predictions about the amount of CRs injected at shocks as a function of various properties of the shock including the Mach number, the obliquity of the magnetic field, or the pre-existing amount of CRs \citep{caprioli18}.  There is a large body of evidence for CRs accelerated in the shocked-shell material of supernova (SN) explosions~\citep[e.g.][]{koyamaetal95,decourchelleetal00,aharonianetal04,warrenetal05, helderetal09, ackermannetal13} and it has been shown that they have a significant impact on the shell structure and dynamics~\citep{chevalier83,dorfi90,zanketal93,wagneretal09, ferrandetal10, castroetal11, pfrommeretal17, paisetal18, diesing18}. Supernova remnants (SNRs) are expected to be the main source of CRs permeating the entire interstellar medium (ISM) of galaxies~\citep{aguilaretal15}, though the consistency of the accelerated CR spectrum in a SNR with that of entire galaxies is still intensely debated~(see \citealp{blasi13} for a review).

Cosmic rays likely  have   an important dynamical impact over the ISM on all galactic scales. On small scales, while released by a SNR, CRs possess enough pressure to overcome the background magnetic and gas pressures and trigger different types of plasma instabilities which result in the production of waves and turbulence~\citep{ptuskin08, malkov13}. This self-generated turbulence can confine CRs over distances and amounts of time that depend on the conditions prevailing in the ISM, especially the ionisation degree~\citep{nava16, nava19}. The generation of waves contribute to locally heating the warm ionised medium~\citep{wiener13}. On larger galactic scales, comparable to the disc  height, CR gradients can modify the dynamics of Jeans unstable regions in the atomic phase~\citep{commerconetal19}, and they can propel cold galactic-wide outflows (\citealp{jubelgasetal08, wadepuhl&springel11, uhligetal12, hanaszetal13, salem&bryan14, salemetal14, girichidisetal16, girichidisetal18, simpsonetal16, recchia17, fujita&maclow18, mao&ostriker18}) with a preferential impact in low-mass galaxies~(\citealp{boothetal13, jacobetal18}, Dashyan \& Dubois, sub.). However, the capability of winds to carry mass and momentum depends on the detailed CR physics such as streaming~\citep{ruszkowskietal17,wieneretal17,holguinetal18,butsky&quinn18}, or taking into account the unresolved multi-phase nature of the gas and its impact on CR transport~\citep{farberetal18}. CRs also boost the dynamo amplification of the magnetic field in disc galaxies~\citep{hanaszetal04,hanaszetal09a,hanaszetal09b,pakmoretal16}. 

On very large cosmological scales CRs are released in shocks~\citep{miniatietal00, miniatietal01, ryuetal03, skillmanetal08, pfrommeretal07, pfrommeretal08, pfrommeretal17, vazzaetal09, vazzaetal12} with external cosmological infall of gas producing the strongest shocks, while pre-processed internal shocks in halos drive the bulk of the shock distribution in the more moderate strength regime.

Similarly, strong shocks are produced in jets from active galactic nuclei;  they release large amounts of CRs as observed in radio emission~\citep{fanaroff&riley74, augeretal07, crostonetal09} and help to release the feedback back to the hot gas from galaxy clusters~\citep{crostonetal08, guooh08, sijackietal08, guo&mathews11, fujita&ohira11, jacob&pfrommer17, ruszkowskietal17agn, ehlertetal18}. However, again, their impact might significantly differ depending on which CR dynamical processes are modelled and which ignored.

In a previous work~\citep{dubois&commercon16}, we  introduced a numerical model for anisotropic CR diffusion. Here, we extend it by including a model of the CR streaming instability and CR injection at shocks through DSA in the adaptive mesh refinement code {\sc ramses}~\citep{teyssier02}. In another work (Brahimi et al. in prep.) we introduce new diffusive transport for CRs accounting for the generation of turbulence produced by the streaming. This ensemble of work aims to provide a consistent description of CR dynamical effect on the  interstellar or intergalactic media. In the same view, a recent model has been proposed by \citet{thomas19}. 

In section~\ref{section:numerics}, we introduce the full set of CR magneto-hydrodynamics including the streaming and acceleration terms, whose numerical modelling and tests are respectively tackled in Sections~\ref{section:streaming} and~\ref{section:acceleration}. We finally test CR acceleration and streaming combined in turbulent interstellar medium experiments in section~\ref{section:ism}.

\section{Magneto-hydrodynamics with cosmic rays}
\label{section:numerics}

By taking the energy moment of the Fokker-Planck CR transport equation~\citep{drury&voelk81}, the following set of differential equations to be solved for cosmic-ray magneto-hydrodynamics (CRMHD) of a fluid mixture made of thermal particles and CRs can be obtained:
\begin{eqnarray}
\label{mass} \frac{\partial \rho}{\partial t} &+& \nabla. (\rho \vec{u})= 0 \, ,\\
\label{momentum}\frac{\partial \rho \vec{u}} {\partial t} &+& \nabla. \left(\rho \vec{u}\vec{u}+P_{\rm tot}-\frac{\vec{B}\vec{B}}{4 \pi}\right)= 0\, ,\\
\label{energy}\frac{\partial e} {\partial t} &+& \nabla. \left((e+P_{\rm tot})\vec{u}-\frac{\vec{B} (\vec{B}.\vec{u})}{4 \pi}\right) = \nonumber \\
&-&P_{\rm CR}\nabla.\vec{u} - \nabla.\vec{F}_{\rm CR,d}+ \mathcal L_{\rm rad} \, ,\\
\label{magnetic}\frac{\partial \vec{B}} {\partial t} &-& \nabla \times (\vec{u} \times \vec{B} )=0\, ,\\
\label{cr}\frac{\partial e_{\rm CR}} {\partial t} &+& \nabla. \left(e_{\rm CR} \vec{u} + (e_{\rm CR}+P_{\rm CR})\vec{u}_{\rm st}\right )= \nonumber\\
&-&P_{\rm CR}\nabla.\vec{u} - \nabla.\vec{F}_{\rm CR,d} + \mathcal L_{\rm st} + \mathcal H_{\rm acc} + \mathcal L_{\rm rad,CR}
.\end{eqnarray}
Here $\rho$ is the gas mass density, $\vec{u}$ is the gas velocity, $\vec{u}_{\rm st}$ is the streaming velocity, $\vec B$ is the magnetic field, 
$e=0.5\rho u^2+e_{\rm th}+e_{\rm CR}+B^2/8\pi$
 is the total energy density,  $e_{\rm th}$ is the thermal energy density, and
 $e_{\rm CR}$ is the CR energy density; 
  $P_{\rm tot}=P_{\rm th}+P_{\rm CR}+P_{\rm mag}$ is the sum of thermal $P_{\rm th}=(\gamma-1)e_{\rm th}$, CR $P_{\rm CR}=(\gamma_{\rm CR}-1)e_{\rm cr}$, and magnetic $P_{\rm mag}=0.5B^2/(4\pi)$ pressures, 
  where $\gamma$ and $\gamma_{\rm CR}$ are the adiabatic indexes of the thermal and CR components, respectively.
We note that all energy components $e_{i}$ are energies per unit volume $e_{i}=E_{i}/\Delta x^3$, where $\Delta x$ is the cell size.
The terms on the right-hand side of the equations are treated as source terms with $P_{\rm CR}\nabla.\vec{u}$ the CR pressure work term, 
$\vec{F_{\rm CR,d}}=-D_0 \vec{b}(\vec{b}. \nabla e_{\rm CR})$ the anisotropic diffusion flux term, $D_0$ the diffusion coefficient (usually taken as a constant value for simplicity, but it can also be a function of local MHD quantities), $\vec{b}=\vec{B}/||\vec B||$  the magnetic unity vector, and a total radiative loss term $\mathcal L_{\rm rad}=\mathcal L_{\rm rad,th}+\mathcal L_{\rm rad,CR->th}$ composed of the thermal $\mathcal L_{\rm rad,th}$ and CR $\mathcal L_{\rm rad,CR->th}$ radiative loss terms, where the CR loss term ($\mathcal L_{\rm rad,CR->th}=\mathcal L_{\rm rad,CR}+\mathcal H_{\rm rad,CR->th}$) is the non-conserving sum of radiative losses from cosmic rays $\mathcal L_{\rm rad,CR}$ turning as a heating rate $\mathcal H_{\rm rad,CR->th}$ for the thermal component.
Finally, and this is the core of this paper, we   detail how the streaming instability terms $\nabla. \left((e_{\rm CR}+P_{\rm CR})\vec{u}_{\rm st}\right )$ (advection-diffusion term) and $\mathcal L_{\rm st}$ (heating term), and the CR acceleration at shocks $\mathcal H_{\rm acc}$ are modelled.

We use the {\sc ramses} code detailed in~\cite{teyssier02} to solve these equations with adaptive mesh refinement (AMR).
The full set of equations is solved with the standard MHD solver of {\sc ramses} described in~\cite{fromangetal06}, where the right-hand side terms of equation~(\ref{energy}) are treated separately as source terms. 
The induction equation~(equation \ref{magnetic}) is solved using constrained transport~\citep{teyssieretal06}, which by construction guarantees at all times that $\nabla.\vec{B}\simeq 0$ at machine precision.
Godunov fluxes are solved with the approximate Harten--Lax--van Leer Discontinuities (HLLD) Riemann solver~\citep{miyoshi&kusano05} and the minmod total variation diminishing slope limiter are modified to account for the extra energy components and total pressure made of the thermal and CR component.
Accordingly, the effective sound speed used for the Courant-Friedrichs-Lewy time-step condition accounts for the extra pressure components (i.e. total pressure of the fluid).
The implementation of the anisotropic CR diffusion in {\sc ramses},  which our new implementation of CR streaming relies on, is described in~\cite{dubois&commercon16}.

It should be noted that equation~(\ref{cr}) can be expanded to as many CR energy bins as required   to sample a full spectrum of CRs in energy-momentum space with source terms communicating the energy fluxes between the various energy bins~\citep[see][for such efforts in those directions]{miniati01, girichidisetal14, winneretal19}.
We ignore this extra level of complexity to represent the entire spectrum of CR energy by a single bin of energy.
For sake of completeness, we introduced the anisotropic diffusion term as well as the CR radiative loss terms~\citep[trivially modelled as a simple density and CR energy-dependent term; see e.g.][]{ensslinetal07, guooh08}  in the equations; we do not make use of them in the various tests of this paper, i.e. $D_0=0$ and $\mathcal{L}_{\rm rad,CR}=0$.

\section{Cosmic-ray streaming}
\label{section:streaming}

\subsection{Numerical implementation}

Cosmic rays propagating faster than the Alfv\'en velocity $\vec{u}_{\rm A}=\vec{B}/\sqrt{4\pi \rho}$ excite Alfv\'en waves, which in turn drive the scattering of the CR pitch angle with magnetic field lines. This coupling leads to a reduced CR bulk velocity at the Alfv\'en velocity and confines the CR streaming transport along the field lines and their own gradient of pressure~\citep{wentzel68, kulsrud&pearce69, skilling75}.
Several damping mechanisms, such as ion-neutral damping, non-linear Landau damping, or turbulence damping~\citep{kulsrud&pearce69, yan&lazarian02,farmer&goldreich04,lazarian&beresnyak06,wieneretal13}, can lead to a significant suppression of these self-excited Aflv\'en waves and can increase the effective value at which CRs are allowed to stream down their own gradient at super-Alfv\'enic velocities $\vec{u}_{\rm st}=-f_{\rm SA}\vec{u}_{\rm A} {\rm sign}(\vec{b}.\nabla e_{\rm CR})$, where   $f_{\rm SA}\ge 1$ is the super-Aflv\'enic boost factor of the streaming velocity.

In addition, while CRs scatter onto the Aflv\'en waves, they experience a drag force, whose work is transferred to the thermal pool at the following rate: 
\begin{equation}
\mathcal L_{\rm st}=-{\rm sign}(\vec{b}.\nabla e_{\rm CR})\vec{u}_{\rm A} .\nabla P_{\rm CR}\, .
\end{equation}
We note that this heating term has $f_{\rm SA}=1$ since only the Alfv\'en waves mediate the energy exchange between CRs and the thermal component~\citep[see e.g.][]{ruszkowskietal17}.
This term, which is by construction always a heating (resp. loss) term for the thermal (resp. CR) component, is obtained by simply differentiating the values of the CR energy density with neighbouring cells.

For simplicity, in the rest of this work, whose aim is to   test the implementation of CR streaming, we  systematically assume $f_{\rm SA}=1$.
The advection or diffusion term of streaming $\nabla.((e_{\rm CR}+P_{\rm CR})\vec{u}_{\rm st})$ can be solved via two distinct approaches. One is to update the CR energy density using an explicit upwind method; however, since the streaming velocity can become discontinuous at extrema of $e_{\rm CR}$, it modifies the condition of stability of the solution to $\Delta t \propto \Delta x^3$~\citep{sharmaetal09}. \cite{sharmaetal09} proposed  regularising the streaming velocity by replacing ${\rm sign}(\vec{b}.\nabla e_{\rm CR})$  by ${\rm tanh}(h\vec{b}.\nabla e_{\rm CR}/e_{\rm CR})$ in order to obtain a less constraining  time-step condition of $\Delta t =h \Delta x^2/(2 e_{\rm CR}u_{\rm A})$, and where $h$ should be the size of a few cells.
Nonetheless, this time-step condition is still too constraining due to the quadratic dependency on cell size, and it is necessary  to rely on a different strategy  to make such a numerical implementation practicable in all possible situations.  \cite{sharmaetal09} suggested using an implicit solver for the regularised upwind method. 
Here we decided to take a different route that relies on the modelling of the anisotropic diffusion with an implicit solver, as done in~\cite{dubois&commercon16}.

We can rewrite the streaming velocity as
\begin{equation}
\vec{u}_{\rm st}= - \frac{\vec{b}.\nabla e_{\rm CR}}{\vert \vec{b}.\nabla e_{\rm CR} \vert} \vec{u}_{\rm A}  \, ,
\end{equation}
which, when recast into $\nabla.((e_{\rm CR}+P_{\rm CR})\vec{u}_{\rm st})$, can be rewritten as a diffusion term~\citep[see also][where the same diffusion approach for the isotropic version of CR streaming is used]{uhligetal12}:
\begin{eqnarray}
 \nabla. \vec{F}_{\rm CR,s}&=& \nabla .(-D_{\rm st} \vec{b}(\vec{b}. \nabla e_{\rm CR})) \nonumber\\
 &=&\nabla . \left( -\frac{ (e_{\rm CR}+P_{\rm CR})\vert B\vert}{\vert \vec{b}.\nabla e_{\rm CR} \vert \sqrt{4\pi \rho}} \vec{b} (\vec{b}. \nabla e_{\rm CR})\right)\, .
\end{eqnarray}
Therefore, this advection-diffusion part of the streaming instability can be treated as an addition to the standard $F_{\rm CR,d}$ CR diffusion term  ($F_{\rm CR, ds}=F_{\rm CR, d}+F_{\rm CR, s}$),  for clarity hereafter written as follows:
\begin{equation}
\nabla.\vec{F}_{\rm CR,ds}= \nabla .\left(-D \vec{b}(\vec{b}. \nabla e_{\rm CR})\right)\, ,
\end{equation}
where $D=D_0+D_{\rm st}$.
The $F_{\rm CR,ds}$ diffusion flux can be arbitrarily decomposed into an anisotropic and isotropic part
\begin{equation}
\nabla.\vec{F}_{\rm CR,ds}= \nabla .\left(-D_\parallel \vec{b}(\vec{b}. \nabla e_{\rm CR})-D_{\rm iso} \nabla e_{\rm CR}\right)\, ,
\end{equation}
where $D_\parallel=(1-f_{\rm iso})D$, $D_{\rm iso}=f_{\rm iso}D$, and $f_{\rm iso}\le1$.
We briefly recall the framework of the implicit solver developed in~\cite{dubois&commercon16}. For the 2D case, the time update of the CR energy by the anisotropic part (the isotropic part is trivially obtained) of the diffusion flux is 
\begin{equation}
\label{eq:2dcond}e^{n+1}_{i,j}+\Delta t \frac{F^{n+1}_{i+\frac{1}{2},j}+F^{n+1}_{i,j+\frac{1}{2}} - F^{n+1}_{i-\frac{1}{2},j} - F^{n+1}_{i,j-\frac{1}{2}}} {\Delta x}= e^{n}_{i,j}\, ,
\end{equation}
where the cell-centred  fluxes are computed with cell-cornered values using the symmetric scheme from~\cite{gunteretal05}:
\begin{eqnarray}
F^{\rm ani}_{i+\frac{1}{2},j}&=&\frac {F^{\rm ani}_{i+\frac{1}{2},j-\frac{1}{2}} + F^{\rm ani}_{i+\frac{1}{2},j+\frac{1}{2}}} {2}\, , \nonumber \\
F^{\rm ani}_{i,j+\frac{1}{2}}&=&\frac {F^{\rm ani}_{i-\frac{1}{2},j+\frac{1}{2}} + F^{\rm ani}_{i+\frac{1}{2},j+\frac{1}{2}}} {2}\, . \nonumber 
\end{eqnarray}
The anisotropic cell corner flux is 
\begin{equation}
\label{eq:aniflux}F^{\rm ani}_{i+\frac{1}{2},j+\frac{1}{2}}=-\bar D \bar b_{x} \left( \bar b_{x} \bar {\frac{\partial e} {\partial x}} + \bar b_{y} \bar{ \frac{ \partial e} {\partial y} }\right)\, ,
\end{equation}
where barred quantities are arithmetic averages over the cells connected to the corner, i.e.
\begin{eqnarray}
\bar b_{x} &=& \frac {b^n_{x,i+\frac{1}{2},j}+b^n_{x,i+\frac{1}{2},j+1}} {2} \, , \nonumber \\
\bar b_{y} &=& \frac {b^n_{y,i,j+\frac{1}{2}}+b^n_{y,i+1,j+\frac{1}{2}}} {2} \, , \nonumber \\
\bar {\frac {\partial e} {\partial x}} &=& \frac {e^{n+1}_{i+1,j+1} + e^{n+1}_{i+1,j} -e^{n+1}_{i,j+1} -e^{n+1}_{i,j}} {2\Delta x} \, , \nonumber \\
\bar {\frac {\partial e} {\partial y}} &=& \frac {e^{n+1}_{i+1,j+1} + e^{n+1}_{i,j+1} -e^{n+1}_{i+1,j} -e^{n+1}_{i,j}} {2\Delta x} \, , \nonumber \\
\bar D &=& \frac{D^n_{i,j}+D^n_{i+1,j}+D^n_{i,j+1}+D^n_{i+1,j+1}} {4} \, .\nonumber 
\end{eqnarray}
We note that all hydrodynamical variables in {\sc ramses} are cell-centred except for the magnetic field which is face-centred.
The streaming diffusion coefficient is computed as 
\begin{equation}
D^n_{i,j}=\frac { (e^n_{i,j}+P^n_{i,j}) } {\sqrt{4 \pi \rho_{i,j}}}  \frac{\vert \widetilde B \vert^n_{i,j}}{\widetilde{\vert \vec{b} . \nabla e \vert}^n_{i,j}} \, , 
\end{equation}
where upper tilde quantities stand for cell-centred quantities reconstructed from a combination of cell-centred and face-centred quantities:
\begin{eqnarray}
\vert \widetilde B \vert^n_{i,j} &=& \frac{1}{2} \sqrt{ \left( B^n_{x,i-\frac{1}{2},j} + B^n_{x, i+\frac{1}{2},j} \right)^2+ \left( B^n_{y, i,j-\frac{1}{2}} + B^n _{y, i,j+\frac{1}{2}}\right)^2 } \, , \nonumber \\
\widetilde{\vert \vec{b} . \nabla e \vert}^n_{i,j} &=& \frac{1}{4 \Delta x \vert \widetilde B \vert^n_{i,j}} \left \vert \left ( B^n_{x,i-\frac{1}{2},j} + B^n_{x, i+\frac{1}{2},j} \right ) \left ( e^n_{x,i+1,j} - e^n_{x, i-1,j} \right ) \right. \nonumber \\
&+&\left. \left ( B^n_{y,i,j-\frac{1}{2}} + B^n_{y, i,j+\frac{1}{2}} \right ) \left ( e^n_{y,i,j+1} - e^n_{y, i,j-1} \right ) \right \vert \, .
\end{eqnarray}
It should be noted that, in principle, the solver can deal with any arbitrary large values of the diffusion coefficient; however, the number of iterative steps of the implicit solver to converge towards the solution can be large for a large diffusion coefficient, typically at  extrema of $\vert \vec{b}.\nabla e_{\rm CR}\vert$ where this value can become close to zero. In practice, we cap the value of the streaming diffusion coefficient to $10^{28} \rm \, cm^2\, s^{-1}$ in all practical astrophysical applications to reduce the spectral condition number of the matrix involved in the implicit solver in order to save  computational iterations.
From the 2D  case, the method is trivially expanded into three dimensions.

\subsection{Tests of CR streaming}

\begin{figure}
\centering \includegraphics[width=0.45\textwidth]{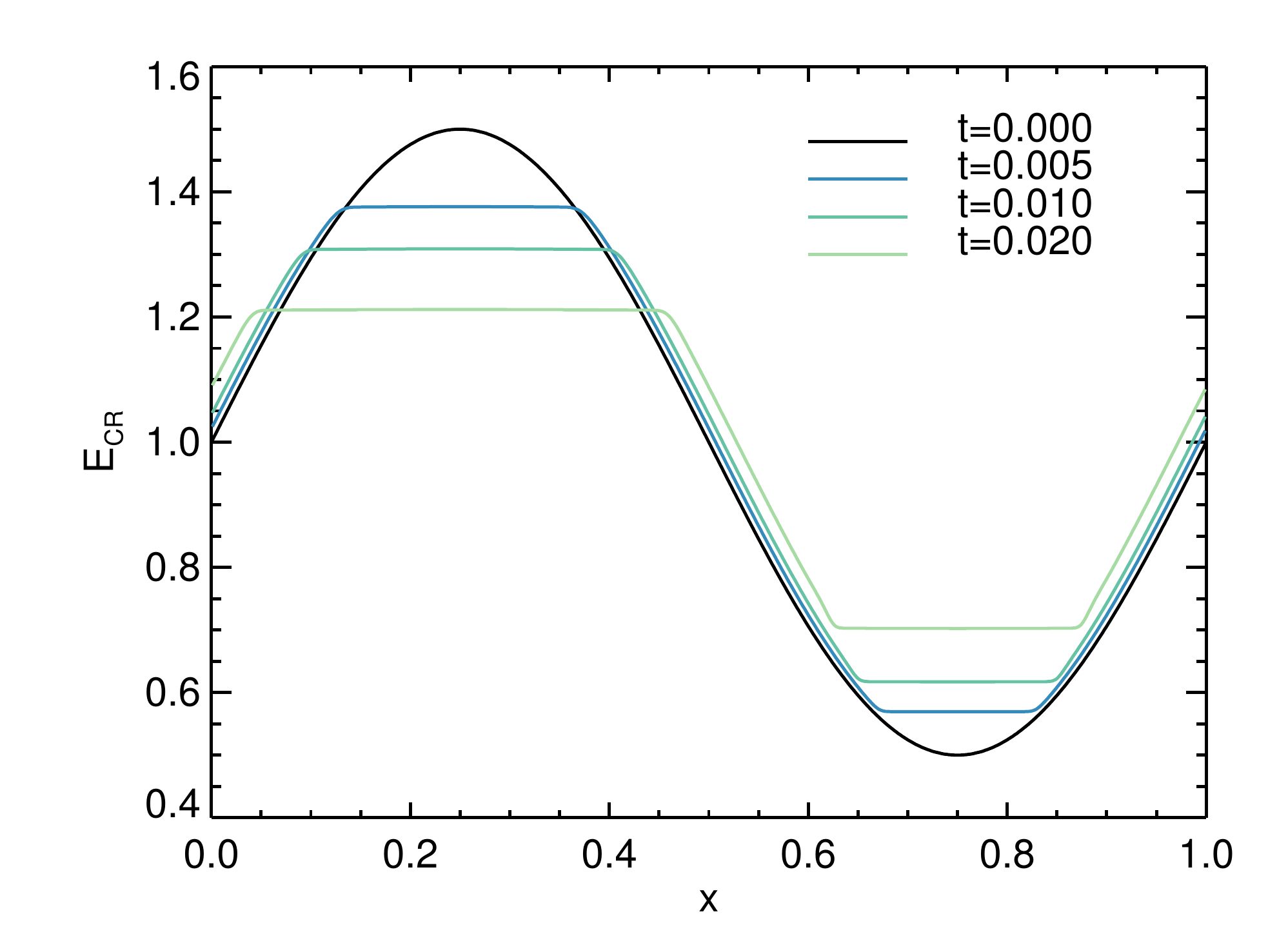}
\caption{Evolution of a 1D sinusoid of CR energy density with streaming advection only as a function of position with 512 cells and imposing a constant Alfv\'en velocity of 1. The solution is made of two plateaus as the maxima are capped over time, due to the infinite streaming diffusion coefficient, while the two regions between the two plateaus move at a velocity of $\pm\gamma_{\rm CR}=\pm1.4$. }
\label{fig:sinusoid}
\end{figure}

\begin{figure}
\centering \includegraphics[width=0.45\textwidth]{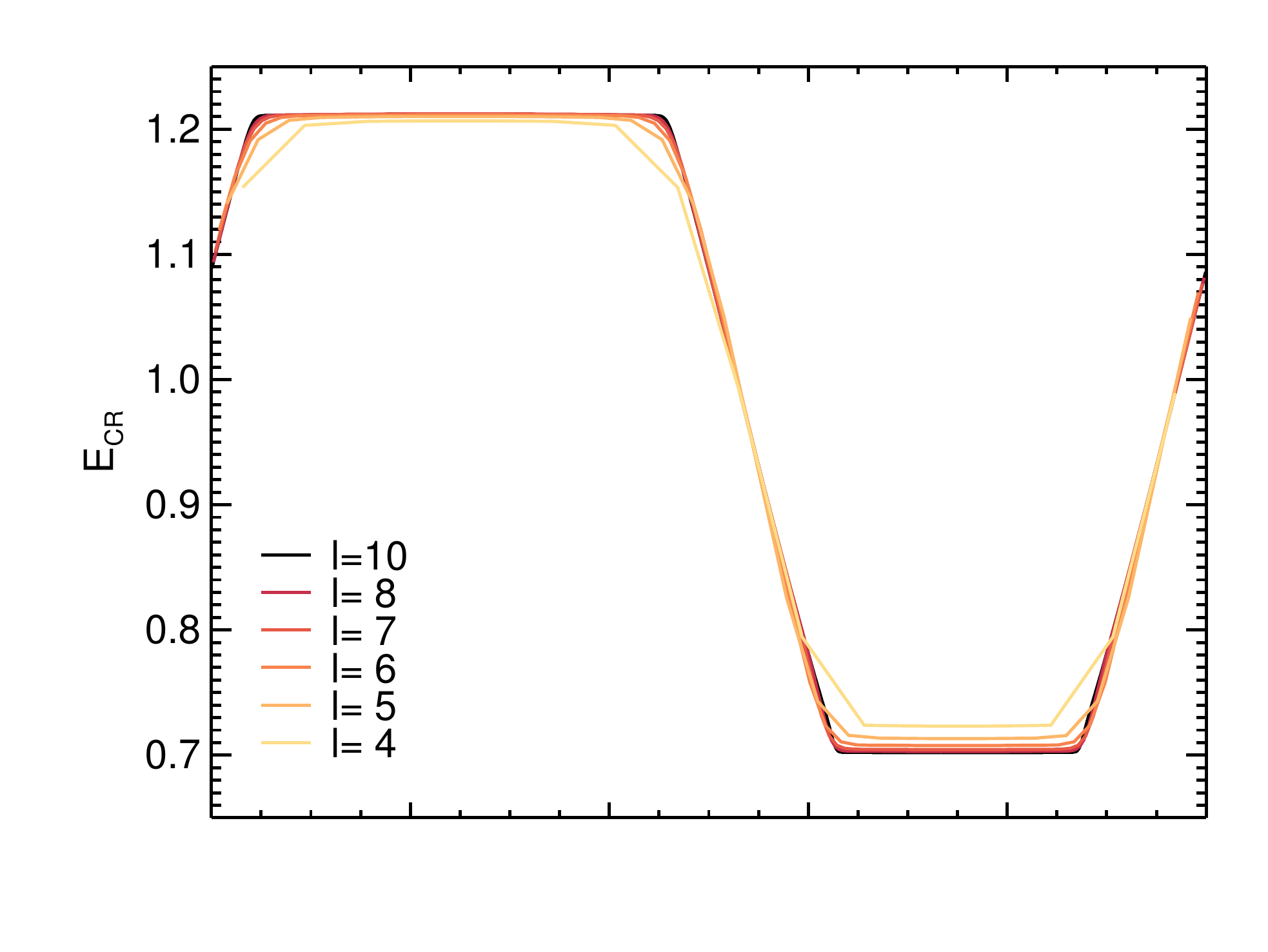}\vspace{-1.3cm}
\centering\includegraphics[width=0.45\textwidth]{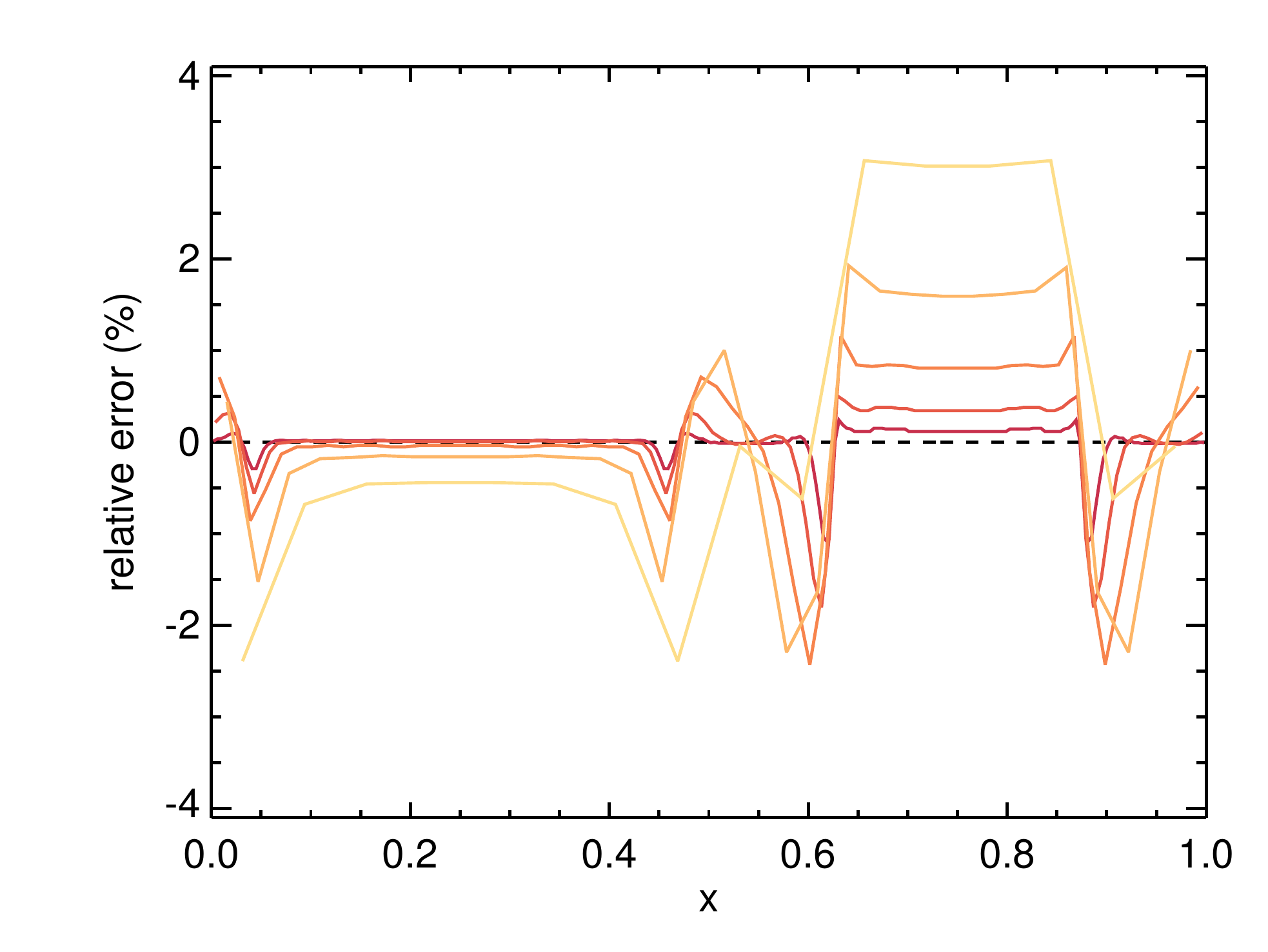}
\caption{Solution at $t=0.02$ of the sinusoid experiment with different resolution from uniform level 4 to 8 (from light red to dark red) and level 10 (in black). The relative errors are compared to the reference numerical solution of level 10. Even for very low resolution the relative error is never larger than a few percentage points. }
\label{fig:sinusoid_ref}
\end{figure}

\begin{figure}
\centering \includegraphics[width=0.45\textwidth]{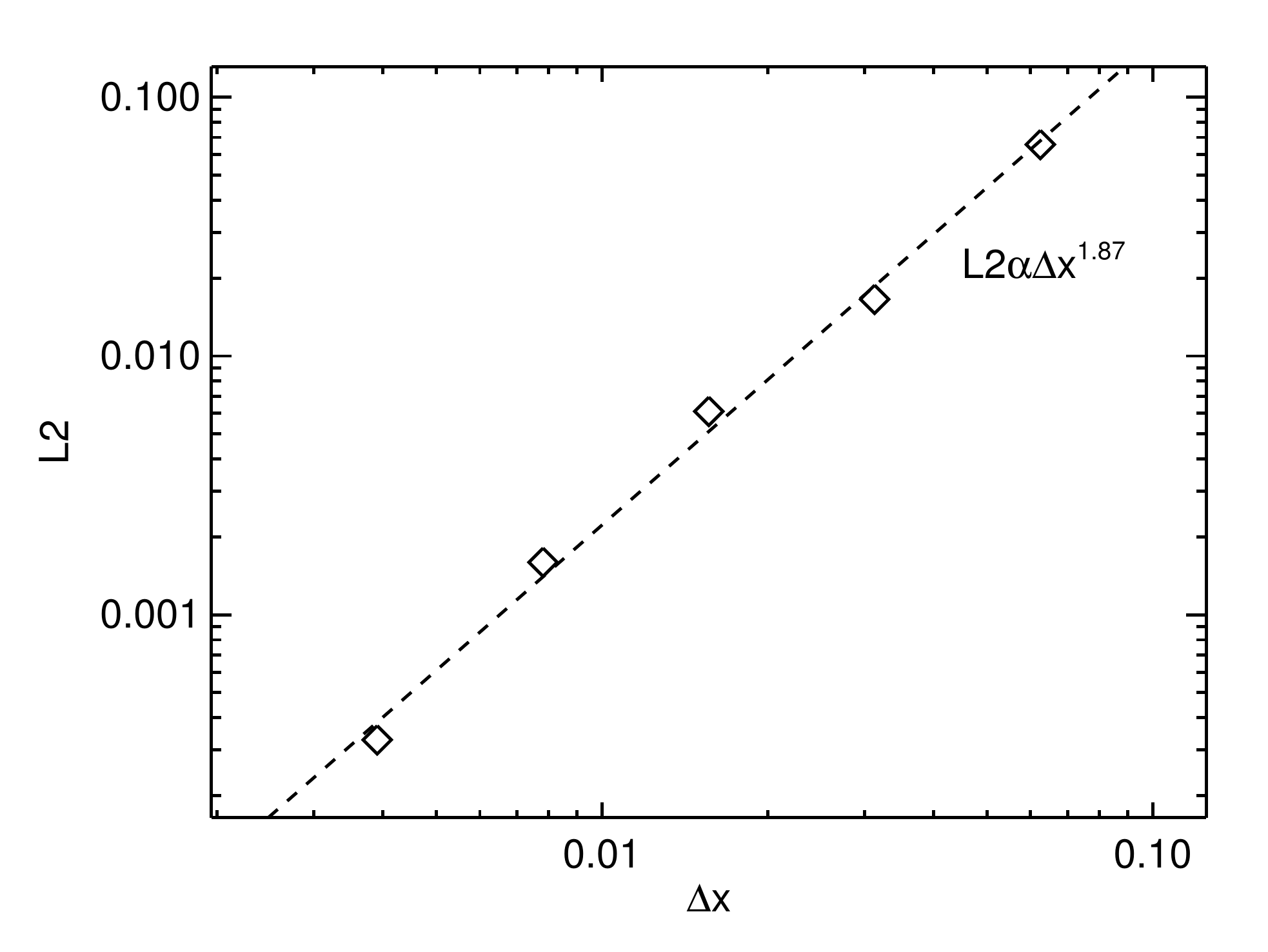}
\caption{Convergence of the L2 norm for the sinusoid experiment using the solution at $t=0.02$. The norm is compared to the reference numerical solution of level 10. The L2 norm scales with $\Delta x^{1.87\pm0.08}$ as indicated by the dashed line.}
\label{fig:sinusoid_lnorm}
\end{figure}

\subsubsection{One-dimensional sinusoid}

In order to test the implementation of the CR advection-diffusion streaming term, a 1D sinusoid experiment is set up where the rest of the physics is deactivated, and with $\gamma_{\rm CR}=1.4$ similar to the test proposed by~\cite{sharmaetal09}.
Unfortunately, there is no known analytical solution to that experiment, but we can test the numerical convergence of the implementation to test its self-consistency.
The initial condition for CR energy density is $e_{\rm CR}=1+0.5\sin(2\pi x)$, and we assume that the Alfv\'en velocity equals 1 oriented along the x-axis.
 In this 1D test we set the maximum streaming diffusion coefficient to be no larger than $100$.
As shown in Fig.~\ref{fig:sinusoid} for this 1D test problem using 512 cells (level 9), the evolved solution is a sinusoid where the  extrema are cropped and where the regions of maximum slope are advected at $\gamma_{\rm CR}\vec{u}_{\rm st}$ (i.e. $-1.4$ if $\partial E_{\rm CR}/\partial x>0$ and $+1.4$ if $\partial E_{\rm CR}/\partial x<0$).
A more evolved time shows a higher cropped fraction of the high and low part of the sinusoid.
We perform a consistency test by varying the resolution of the simulation from 16 cells to 1024 cells, where the highest resolution simulation is used as a reference for comparison.
Figure~\ref{fig:sinusoid_ref} shows the solution at time $t=0.02$ for  16, 32, 64, 128, 256, and 1024 cells, and their relative variation to the reference run.
The solution shows very good numerical convergence towards the high-resolution reference solution, which never exceeds a few percentage points relative variation even when using only 16 cells to resolve the wavelength of the sinusoid.
Finally, the L2 norm (again using the 1024-cell run as a reference) is computed and has a
 convergence with a scaling of $\Delta x^{1.87\pm0.08}$, as shown in Fig.~\ref{fig:sinusoid_lnorm}.

\subsubsection{Two-dimensional sinusoid in a looped magnetic field}
\label{section:sinusoid2d}

\begin{figure}
\centering \includegraphics[width=0.24\textwidth]{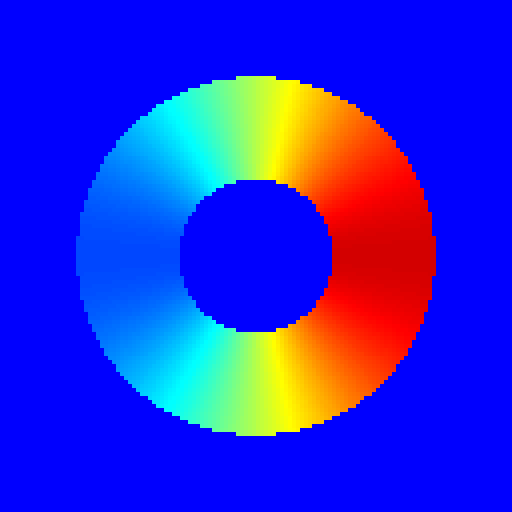}
\centering \includegraphics[width=0.24\textwidth]{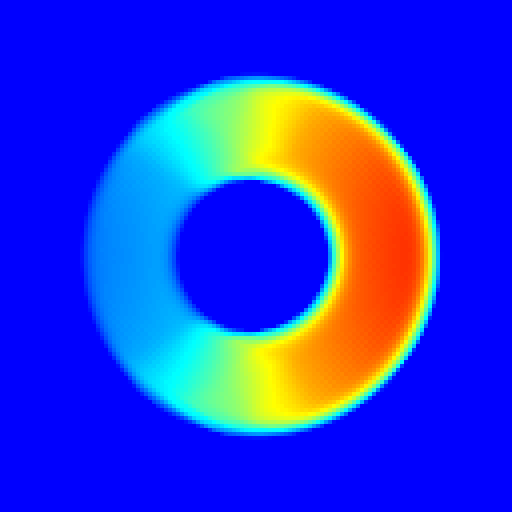}
\centering \includegraphics[width=0.5\textwidth]{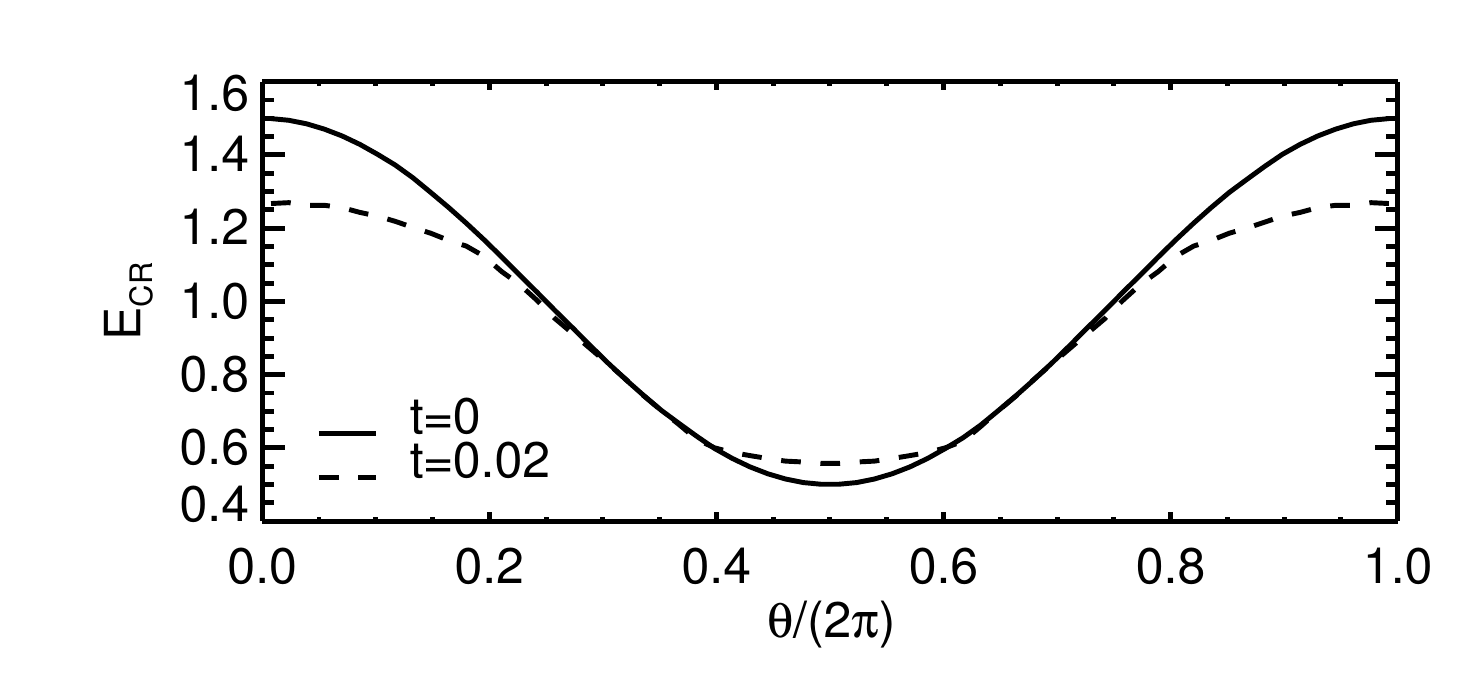}
\vspace{-0.5cm}
\caption{Cosmic-ray energy density maps at $t=0$ (top left) and $t=0.02$ (top right) for an initial angle-dependent sinusoid within a purely circular magnetic field with an Alfv\'en velocity of 1 and a resolution of $128^2$ cells. The energy is evolved with the streaming advection-diffusion term only. The bottom panel shows the radially averaged energy in the radius interval $r=[0.15,0.35]$ as a function of the polar angle $\theta$.}
\label{fig:loop_sinusoid}
\end{figure}

In this test case we try to mimic the 1D sinusoid problem embedded in a non-uniform magnetic configuration.
We initialise a 2D looped magnetic field centred on the middle of the box, hence in the circular coordinate system the magnetic field is purely tangential.
We also initialise the CR energy density in the same way as  the previous 1D test case with a $\theta$ angle dependency $e_{\rm CR}=1+0.5\sin(\theta)$ for a radius $0.15<r<0.35$ and $e_{\rm CR}=10^{-5}$ for $r\le 0.15$ and $r\ge 0.35$.
In this 2D test we set the maximum streaming diffusion coefficient to be no larger than $1$ and an isotropic component of $f_{\rm iso}=10^{-2}$; we discuss the effect of changing these values  on the solution in Appendix~\ref{appendix:sinusoid2d_kiso}.
We choose an Alfv\'en velocity of 1, and again we deactivate the rest of the hydrodynamics.
Figure~\ref{fig:loop_sinusoid} shows the result at times $t=0$ and $t=0.02$.
The solution shows a similar angle-dependent pattern for the evolved solution at $t=0.02$ to that of the 1D case at the same time (i.e. the value of energy density is close to uniform around regions of initial extrema).
We note that the capping of extrema is slightly late in this 2D configuration with respect to the  1D test: compared with Fig.~\ref{fig:sinusoid}, where the maximum and minimum are respectively 1.2 and 0.7 at time $t=0.02$, here in 2D we obtain 1.28 and 0.6, respectively.
We  also tested the 2D streaming for a  ten times wider range of initial CR energy density.
The result, not shown here, is qualitatively similar to that of our reference test.

\section{Shock-accelerated CRs}
\label{section:acceleration}

\subsection{Shock finder algorithm}

Our shock finder algorithm relies on several criteria.
A shock cell is identified as such when all of the following conditions are met: i) $\nabla T . \nabla S>0$ (\citealp{ryuetal03}, where $S=T/n^{2/3}$ is the pseudo-entropy) and $\nabla T . \nabla \rho>0$~\citep[which filters out tangential discontinuities,][]{schaal&springel15};  ii) $\nabla . \vec{u}$ is negative (compression region); iii) $\nabla.\vec{u}$ is a local minimum along the normal to $\vec{n}_{\rm s}=-\nabla T/\vert \nabla T \vert$ (where the local value of $\nabla.\vec{u}$ is compared to the cloud-in-cell interpolated value of $\nabla . \vec{u}$ at one $\Delta x$ local cell distance in the upstream and downstream of the local cell); and  iv) the Mach number is larger $\mathcal M> \mathcal M_{\rm min}$, with $\mathcal M_{\rm min}\simeq 1.5$.
Keeping in mind these conditions,   the Mach number of eligible cells is computed according to the criteria using upstream (pre-shock) and downstream (post-shock) fluid variables.
Using the Rankine-Hugoniot shock jump relations, the Mach number can be computed from density, temperature, or   pressure values. 
For instance, the Mach number for a single thermal component can be obtained from the ratio $\mathcal{R}_{\rm P}=P_{\rm 2}/P_{\rm 1}$ of the downstream to upstream pressures (here and in the following we keep the 1 and 2 subscripts for the upstream and downstream quantities), leading to
\begin{equation}
\label{eq:mach}
\mathcal M^2=\frac{1}{2\gamma}\left[(\gamma-1)+(\gamma+1)\mathcal{R_{\rm P}} \right]\, .
\end{equation}
We note that it is  also possible to employ the jump relations for density or velocity; however, they quickly saturate at high Mach numbers, while pressure jumps offer better leverage for probing the values of the Mach number.

Since our aim is to apply this shock finder to a thermal--CR mixture, the following relation~\citep{pfrommeretal17} should be used instead:
\begin{equation}
\mathcal M^2=\frac{1}{\gamma_{\rm e}} \frac{\mathcal{R}_{\rm P}\mathcal{C}}{\mathcal{C}-\left[(\gamma_1+1)+(\gamma_1-1)\mathcal{R}_{\rm P}\right](\gamma_2-1)} \, .
\end{equation}
Here $\mathcal{C}=[(\gamma_2+1)\mathcal{R}_{\rm P}+(\gamma_2-1)](\gamma_1-1)$, $\gamma_i=P_i/\epsilon_i+1$ for $i=\{1,2\}$ (respectively upstream and downstream) and $\gamma_{\rm e}= (\gamma P_{\rm th,2}+ \gamma_{\rm CR} P_{\rm CR,2})/P_2$ for the downstream region.
In the limit where the weighted adiabatic indexes are equal $\gamma_{\rm e}=\gamma_1=\gamma_2$ this formula for the Mach number is equal to the classical formulation of equation~(\ref{eq:mach}).

The normal to the shock is provided by the gradient of temperature $\vec{n}_{\rm s}$. 
A first guess of the upstream and downstream values of pressure are obtained by cloud-in-cell interpolating the values of the $2^{D}$ cell pressure (where $D$ is the dimensionality of the system to simulate), one cell and two cells away from the shocked cell candidates along $\vec{n}_{\rm s}$ and $-\vec{n}_{\rm s}$ for the upstream and downstream quantities, respectively. 
The upstream and downstream pressures are respectively the minimum and maximum of pressures obtained from the one cell and two cell distances away from the shocked cell.
This first guess of the Mach number is kept for cells with moderate Mach numbers $\mathcal{M}<5$, while cells with higher Mach numbers require  probing regions further than two cells away from the shocked cell to properly evaluate their Mach numbers.
As we  see in the tests, the stronger the shock, the larger the number of cells to sample the discontinuity, and we thus need to probe more distant cells to accurately capture the true upstream and downstream values of the shock. 
This first guess is limited to two cells to fully exploit the code structure of {\sc ramses} that tracks at each time the $3^{D}-1$ neighbouring octs  of each cell (an oct contains $2^D$ cells), including virtual octs that belong to another domain   (hence, going further away requires communication between CPU domains and can be prohibitive, which is why we limit this search to the strongest shocked cells).

The second guess of the Mach number, and other related quantities (see next section), is obtained by moving forward along the normal to the shock by steps of $\Delta x$ up to four cells distance, thus probing both $3\Delta x$ and $4\Delta x$ in  the upstream and the downstream regions. 
For the new value of upstream and downstream pressures (and other related quantities) to be accepted for the calculation of the new Mach number, we check that  the slope of the thermal energy is getting shallower (the profile must flatten as we are moving outwards) by computing the new gradient of thermal energy and comparing to its value from the previous distance step, and  that  the total pressure and the density both have a new extremum (either an upstream minimum or a downstream maximum).
Our experiments with Mach numbers as strong as 1000 has lead us to use up to four cells distance to probe the estimated Mach number of strong shocks, hence we  always use this maximum value in the following, but our implementation can work with arbitrarily larger distances.

\subsection{Cosmic-ray acceleration at shocks}
\label{section:cracc}

At shocks the kinetic energy flux of the upstream flow $\phi_{\rm K,1}=0.5\rho_{\rm 1}u_{\rm 1}^3$ (where the velocities are measured in the moving shock frame) is dissipated by the shock interface into a thermal energy flux $\phi_{\rm th,2}=e_{\rm th,diss}u_2$, CR energy $\phi_{\rm CR,2}=e_{\rm CR,diss}u_2$ and the remaining into kinetic and magnetic energy.
For classical strong shocks without CR acceleration, the ratio of post-shock thermal (dissipated) energy to the pre-shock kinetic energy $e_{\rm th,diss}/(0.5\rho u_{1}^2)$ can be obtained from the Rankine-Hugoniot jump relations, and tends towards $0.56$ for $\gamma=5/3$.
Once shocked cells are identified, the amount of accelerated CRs is obtained with the CR flux following
\begin{equation}
\phi_{\rm CR}=\eta(\mathcal M, X_{\rm CR}, \theta_{\rm B}) e_{\rm diss} u_{\rm 2}\, ,
\end{equation}
where $e_{\rm diss}=e_{\rm th,diss}+e_{\rm CR,diss}$ is the dissipated internal energy of the gas, $u_2$ is the downstream velocity in the frame of the moving shock, and $\eta(\mathcal M, X_{\rm CR}, \theta_{\rm B})$ is the acceleration efficiency of CRs at shocks, which is a function of the Mach number, the upstream CR-to-thermal ratio $X_{\rm CR}=P_{\rm CR,1}/P_{\rm th,1}$, and the magnetic obliquity to the normal of the shock $\theta_{\rm B}$.
Instead of measuring the downstream velocity in the shock frame (which requires  knowing both the upstream and downstream velocities in the lab frame, as well as the jump density ratio $\mathcal{R}_\rho$), we replace $u_2$ by $\mathcal{M}c_{\rm s, 1}/\mathcal{R}_\rho$, where $c_{\rm s,1}$ is the upstream sound speed.
The dissipated energy can be directly measured from the upstream and downstream thermal and CR energy densities
\begin{equation}
e_{\rm diss}= e_{\rm th, 2}+e_{\rm CR, 2}-e_{\rm th, 1} \mathcal{R}_\rho^{\gamma}-e_{\rm CR, 1} \mathcal{R}_\rho^{\gamma_{\rm CR}}\, ,
\end{equation}
where $e_{\rm th, 2}$ and $e_{\rm th, 1}$ are respectively the downstream and upstream thermal energy densities, $e_{\rm CR, 2}$ and $e_{\rm CR, 1}$ the downstream and upstream CR energy densities, and $\mathcal{R}_\rho$  the jump density ratio. 
The jump density ratio is obtained from the direct evaluation of the upstream and downstream densities \begin{equation}
\mathcal{R_\rho}= \frac{\rho_2}{\rho_1}\, .
\end{equation}
The $\mathcal{R}_\rho^{\gamma}$ and $\mathcal{R}_\rho^{\gamma_{\rm CR}}$ terms account for the fact that the upstream thermal and CR energies are also adiabatically compressed at the shock.
Finally, the new CR energy is updated using $\Delta e_{\rm CR}=\phi_{\rm CR} \Delta t/\Delta x$.

According to detailed simulations of accelerated CRs at shocks~\citep{caprioli&spitkovski14acc}, their acceleration efficiency depends on both the Mach number of the shock and the upstream magnetic field orientation with respect to the normal to the shock $\theta_{\rm B}=\arccos(\vec{b_1}.\vec{n}_{\rm s})$.
The dependency of the efficiency of CR acceleration with this so-called `magnetic obliquity' can be factorised out,
$\eta(\mathcal{M},X_{\rm CR}, \theta_{\rm B})=\eta_0\xi(\mathcal{M},X_{\rm CR})\zeta(\theta_{\rm B}),$
and approximated by the following functional form~\citep{paisetal18}:
\begin{equation}
\zeta(\theta_{\rm B})= \frac{1}{2} \left[ \tanh \left( \frac{\theta_{\rm crit}-\theta_{\rm B}}{\delta_{\theta}} \right ) +1 \right] \, , 
\end{equation}
where $\theta_{\rm crit}=\pi/4$ and $\delta_{\theta}=\pi/18$.
Therefore, we probe the angle $\theta_{\rm B}$ by evaluating the orientation of the magnetic vector in the upstream region using the cell that defines the value of the upstream pressure as defined in the previous section.

\begin{figure}
\centering \includegraphics[width=0.45\textwidth]{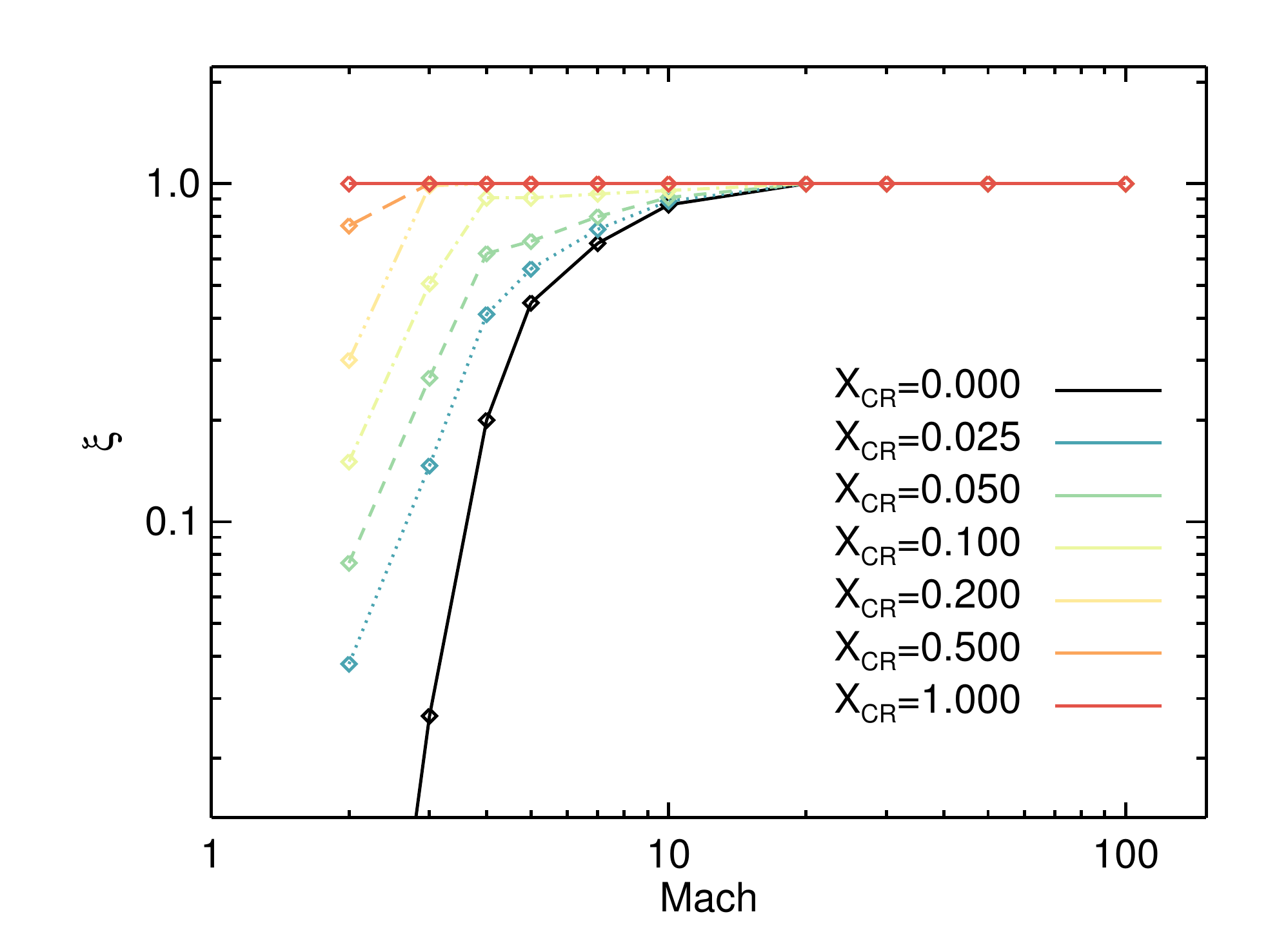}
\caption{Acceleration efficiency $\xi(\mathcal{M},X_{\rm CR})$ as a function of the Mach number $\mathcal{M}$ for different values of the upstream CR-to-thermal pressure ratio $X_{\rm CR}$. The values are obtained from the $X_{\rm CR}=0$ and $0.025$ values of~\cite{kang&ryu13} and renormalised to a maximum value of 1. }
\label{fig:kr13}
\end{figure}

The dependency of the acceleration $\xi(\mathcal{M},X_{\rm CR})$ is obtained from the results of~\cite{kang&ryu13}, and is an increasing function of both $\mathcal{M}$ and $X_{\rm CR}$.
They provide values of the acceleration efficiency for two values of $X_{\rm CR}$, namely $0$ and $0.05$, and ten values of the Mach number (from $1.5$ to $100$). 
Since, to the best of our knowledge, no work has explored the cases with $X_{\rm CR} > 0.05$, in order to   explore the full range of admissible values of $X_{\rm CR}$ we simply interpolate and extrapolate the values of $\xi(\mathcal{M},X_{\rm CR})$ from $X_{\rm CR}=0$ and $0.05$, sampling values of $X_{\rm CR}=0.025$, $0.1$, $0.2$, $0.5$, and $1$.
In addition, we fix those sampling values so that $\xi$ is a monotonic increasing function of $\mathcal{M}$ and $X_{\rm CR}$.
We note that their obtained values of the acceleration efficiency saturates at $\eta_0=0.225$, a factor of $\sim 2$ larger than the maximum values obtained by~\cite{caprioli&spitkovski14acc} for parallel shocks ($\theta_{\rm B}=0$).
We thus renormalise $\xi(\mathcal{M},X_{\rm CR})$ by $0.225$ so that the maximum allowed efficiency is explicitly controlled by $\eta_0$.
The values of $\xi$ are shown in Fig.~\ref{fig:kr13} and are available as tabulated values in Appendix~\ref{app:accKR13}.
We note that obliquity-dependent CR acceleration simulations conducted by~\cite{caprioli18} with a pre-existing population of CRs in the upstream region suggest that the transition of the obliquity-dependent part of the efficiency $\zeta(\theta_{\rm B})$ from the efficient to the inefficient regime is displaced from $\theta_{\rm crit}=\pi/4$ to  $\theta_{\rm crit}=\pi/3$. 
We neglect this effect at the moment.

Finally, we decided to inject the CR energy accelerated at shocks a few cells away from the shock cell. 
We were guided by the fact that numerical shocks are not pure discontinuities and are in fact numerically broadened; therefore, any CR pressure deposited in the numerically broadened shock layer  experiences a work $P_{\rm CR}\nabla.\vec{u}$ of pressure forces. 
For this reason, the CR energy is deposited in the cell of minimum $| \nabla.\vec{u}|$ in the post-shock direction   up to four cells away from the shock cell.
We emphasize that this choice is crucial to obtaining the correct amount of CR energy density in the post-shock region, and our experiments have taught us that the direct injection in the shock systematically overestimates the resulting CR energy density in the post-shock region by a large factor even in the simplest 1D test case (e.g. by a factor of $\sim 2$ for the Sod test).

\begin{figure}
\centering \includegraphics[width=0.5\textwidth]{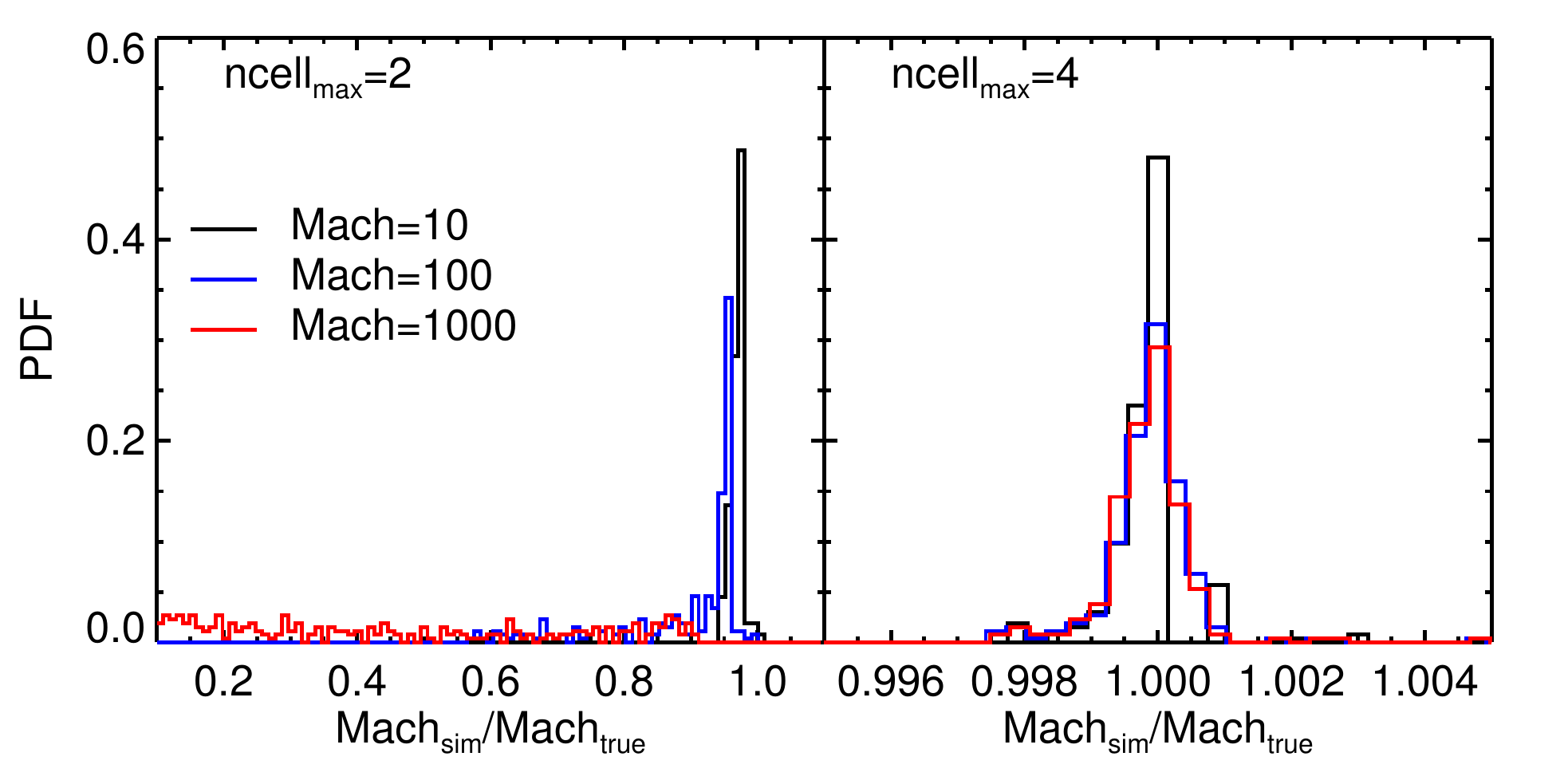}
\caption{Statistics of the numerical Sod shock Mach number relative to its expected value for different shock Mach numbers, 10 (black), 100 (blue), and 1000 (red), using either a maximum of ncell$_{\rm  max}=2$ cells (left panel) or ncell$_{\rm  max}=4$ cells (right panel) to probe hydrodynamical values in the post-shock and pre-shock regions. These shock tube tests do not model CRs. Pre-shock and post-shock regions need to be probed up to four cells away from the shock cell location for the strongest Mach numbers to be captured accurately.}
\label{fig:sod_mach_stat}
\end{figure}

\begin{figure}
\centering \includegraphics[width=0.24\textwidth]{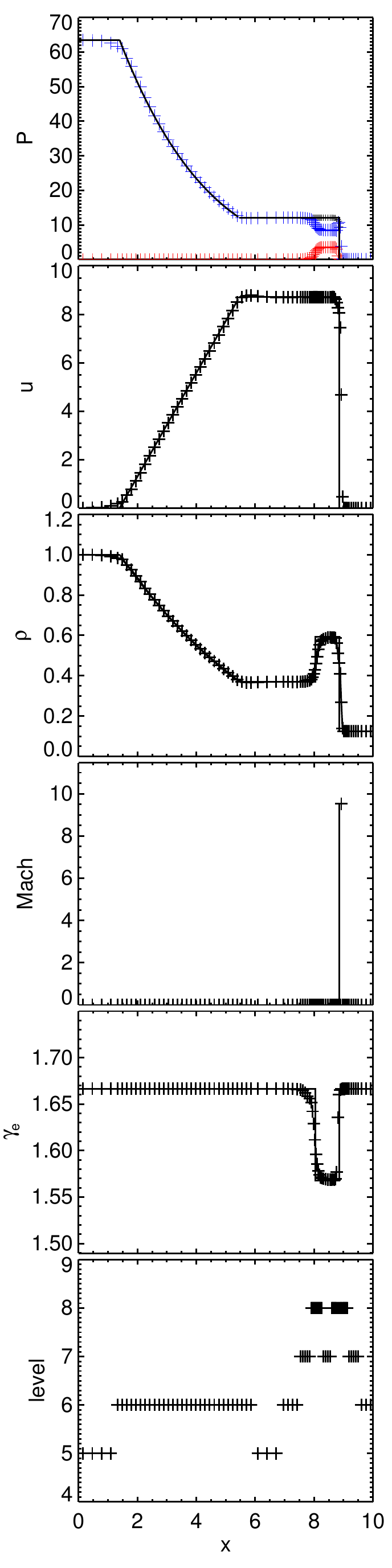}\hspace{-0.25cm}
\centering \includegraphics[width=0.24\textwidth]{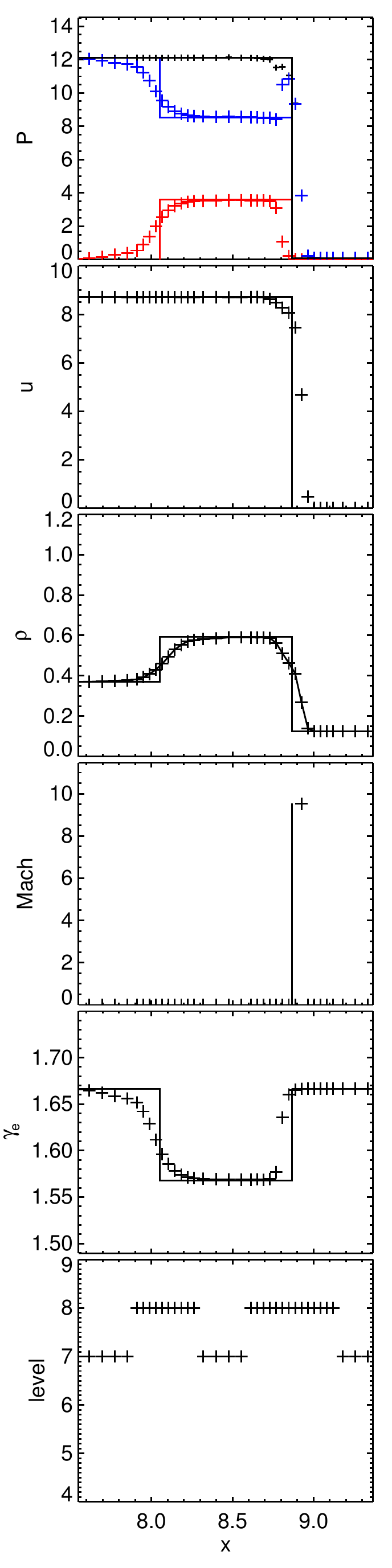}
\caption{Sod shock tube experiment with CR acceleration efficiency of $\eta=0.5$, zero initial CR pressure and $\gamma_{\rm CR}=4/3$ at $t=0.35$. The left panels show the solution over the full box, while the right panels show a zoomed-in region over the shock and contact discontinuities for better clarity of the CR shock-accelerated region. From top to bottom are the pressures (black: total, blue: thermal, red: CR), the density, the velocity, the Mach number, the effective adiabatic index, and the level of refinement. The symbols stand for the numerical solution, while the solid lines are for the analytical solution. The exact Sod solution with accelerated CRs is  reproduced well by our numerical implementation.}
\label{fig:sod}
\end{figure}

\begin{figure}
\centering \includegraphics[width=0.4\textwidth]{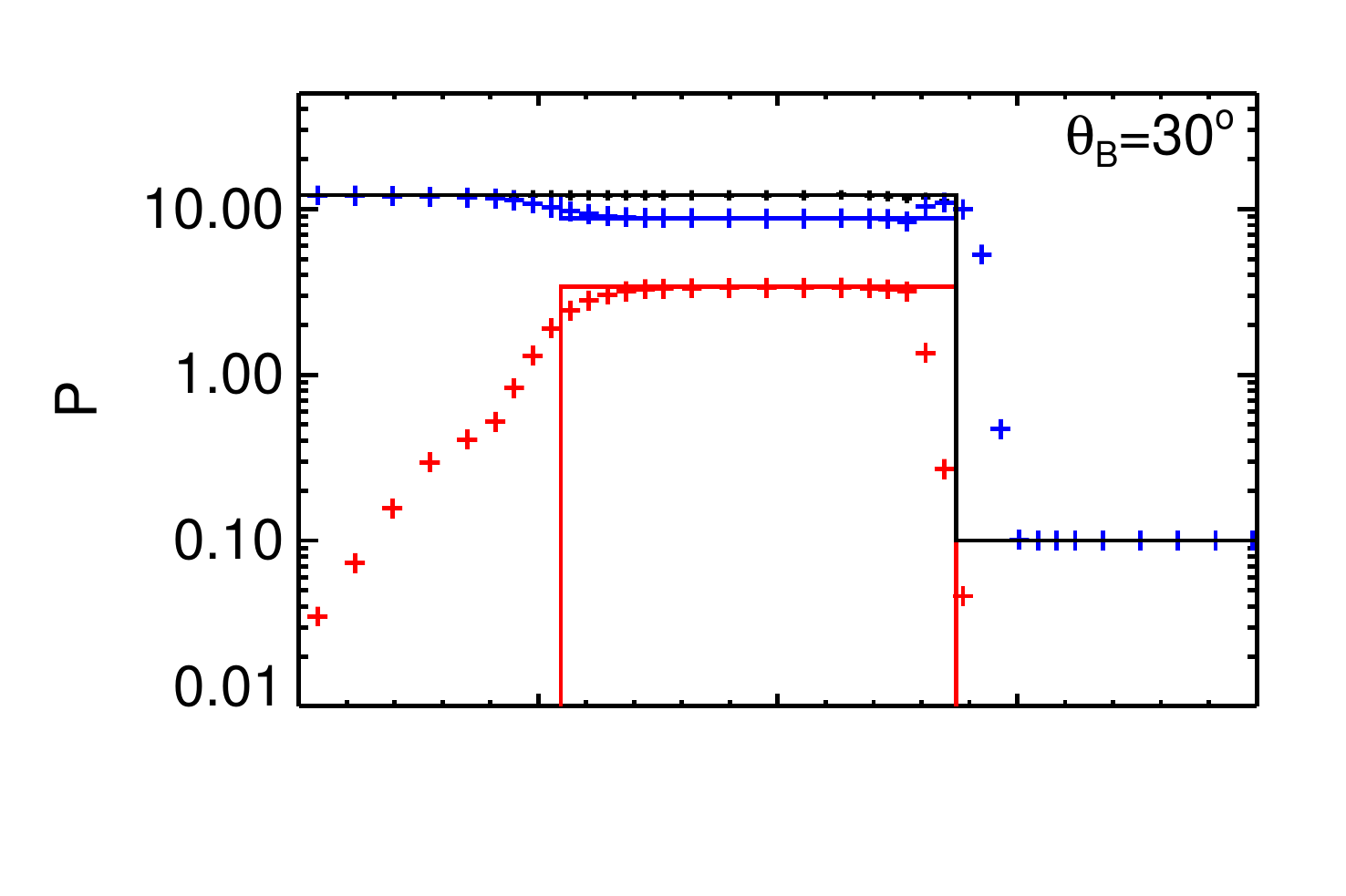}\vspace{-1.5cm}
\centering \includegraphics[width=0.4\textwidth]{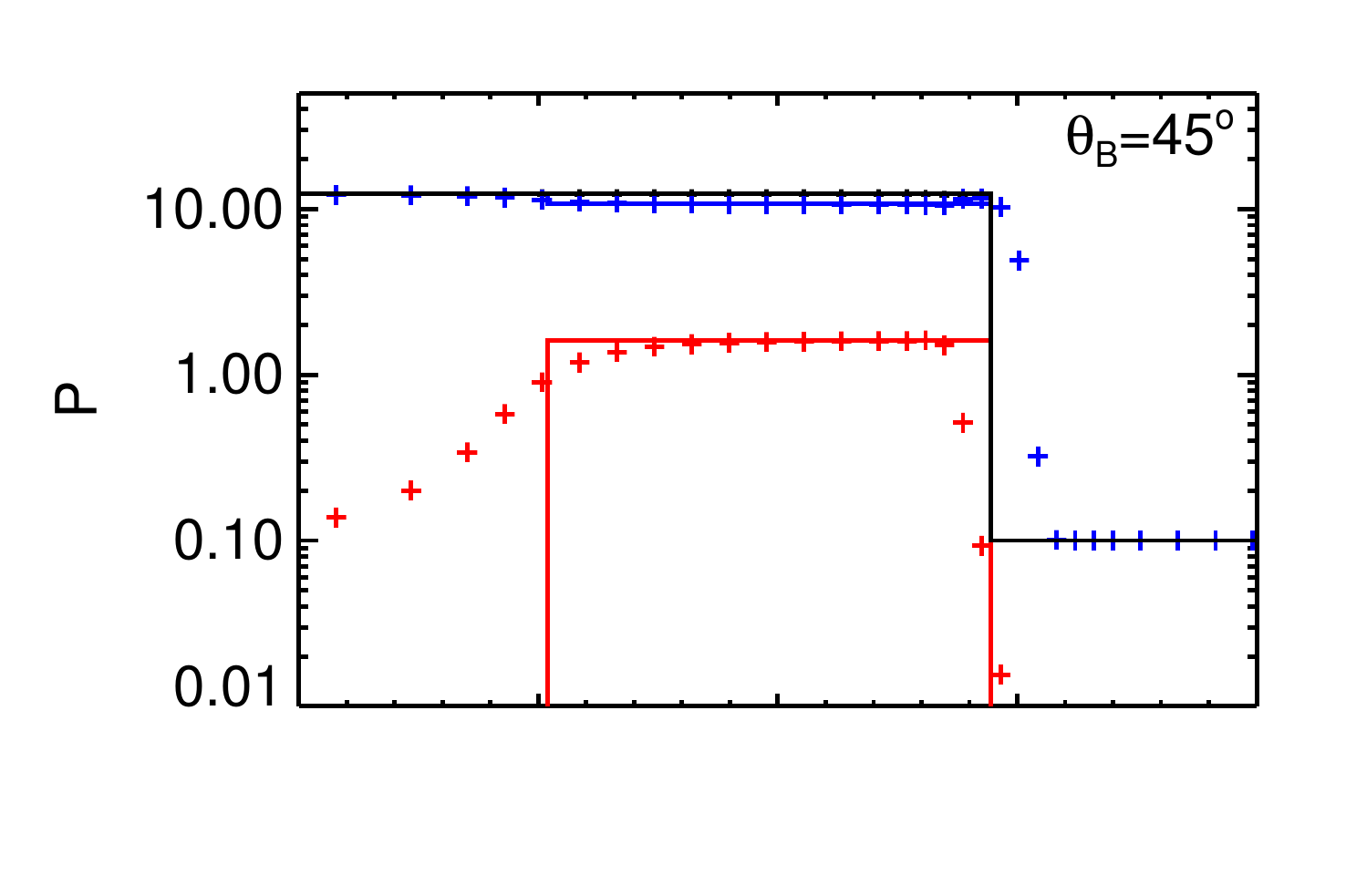}\vspace{-1.5cm}
\centering \includegraphics[width=0.4\textwidth]{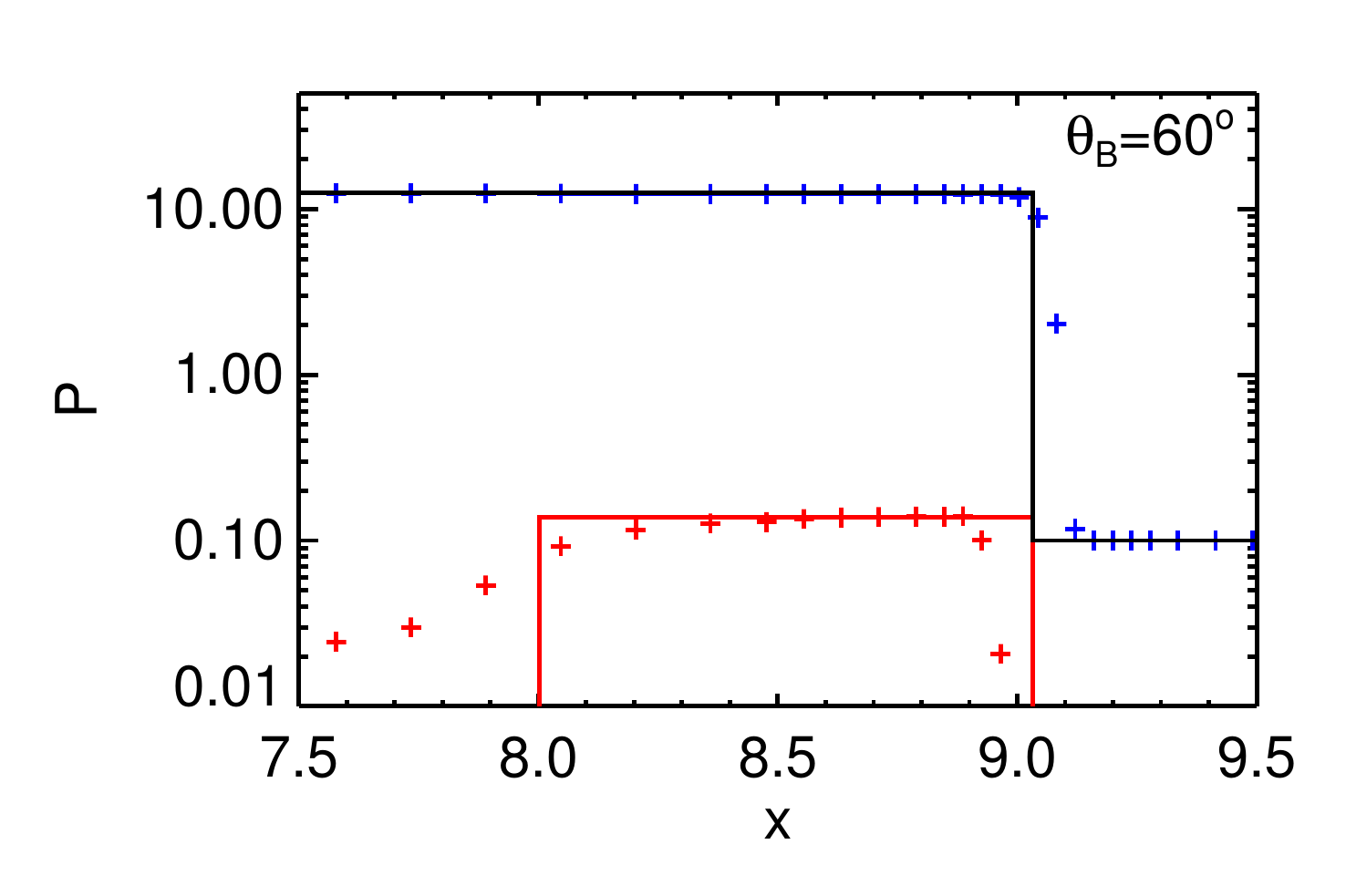}
\caption{Sod shock tube experiment with zero initial CR pressure and $\gamma_{\rm CR}=4/3$ with obliquity-dependent CR acceleration efficiency ($\theta_{\rm B}$ is the so-called obliquity: angle of the pre-shock $\vec{B}$ field with the normal to the shock) $\eta_0=0.5\zeta(\theta_{\rm B})$ for $\theta_{\rm B}=30,45,60^\circ$ from top to bottom. The  panels show the pressures (black: total, blue: thermal, red: CR) at $t=0.35$ over a zoomed-in region over the shock and contact discontinuities for better clarity of the CR shock-accelerated region. The symbols stand for the numerical solution, while the solid lines are for the analytical solution. As expected, the amount of CRs produced at the shock decreases with  obliquity, and reproduces well the exact solution.}
\label{fig:sod_theta}
\end{figure}

\subsection{One-dimensional Sod shock tube}

\subsubsection{Convergence of the shock Mach number}
\label{sec:sodnocr}

In this first test for the convergence of the evaluated shock Mach number, we used the standard Sod shock tube initial conditions for a Mach of 10; in other words, we started with initial left and right states separated by a virtual interface at $x=5$ in a box of size of  $10$ with thermal pressure $P_{\rm th,L}=63.499$ and $P_{\rm th,R}=0.1$, density $\rho_{\rm L}=1$ and $\rho_{\rm R}=0.125$, velocity $u_{\rm L}=u_{\rm R}=0$.
This test was run without any initial or accelerated CR component (i.e. free of CR pressure), and we adopted an adiabatic index of the gas of $5/3$.
In addition we also explored more aggressive shock tube initial conditions to probe Mach of 100 ($P_{\rm th, L}=6349.9$), and Mach of 1000 ($P_{\rm th, L}=634990$).
We employed a base grid of level 5 with up to three additional levels of refinements triggered in regions where the relative cell-to-cell variation of either the density, velocity, or pressure is larger than 10 \%.

Figure~\ref{fig:sod_mach_stat} shows the quality of the Mach number evaluation with the statistics of its value relative to the exact analytical value for various shock tube tests, changing the strength of the shock by two orders of magnitude. 
We tested two maximum values of the extent of the pre-shock and post-shock quantities, either probing up to ncell$_{\rm  max}=2$ cells or ncell$_{\rm  max}=4$ away from the shock cell.
We note that we   removed the estimates of the Mach number for the first 15 time steps of the simulations (over the 263 available time steps, reaching final times $t=0.35$, $t=0.035$, and $t=0.0035$ for Mach numbers of 10, 100, and 1000, respectively), where the shock, contact, and rarefaction waves are not yet sufficiently separated to correctly capture the Mach number of the shock.
It shows that ncell$_{\rm  max}=2$ cells can be sufficient to obtain Mach numbers accurate to a  level of a few  percentage points up to Mach numbers of the order of $\sim100$, even  though it is systematically underevaluated; however, Mach numbers of 1000 are almost never correctly captured.
On the contrary, going up to ncell$_{\rm  max}=4$ cells distance to measure hydrodynamical quantities involved in the reconstruction of the Mach number allows   a precision of better than $0.1$ \%\ in this simple 1D shock tube test.
This behaviour is the natural outcome of the larger numerical broadening of shock discontinuities for stronger shocks (see Appendix~\ref{appendix:shock_broadening}): strong shocks require more cells to resolve the entire shock layer.
We note that increasing the level of refinement does not cure the problem; the  shocks are narrower in physical extent, but   the number of cells required to describe the shock jump remains the sam.

\begin{figure*}
\centering \includegraphics[width=0.33\textwidth]{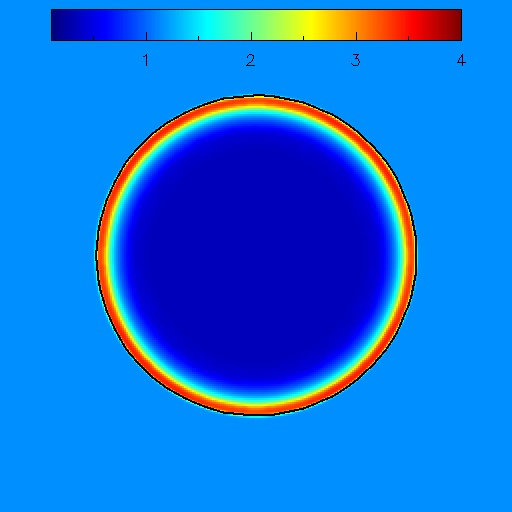}
\centering \includegraphics[width=0.33\textwidth]{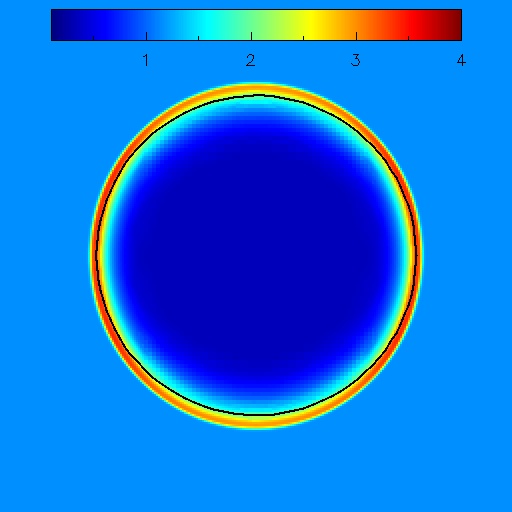}
\centering \includegraphics[width=0.33\textwidth]{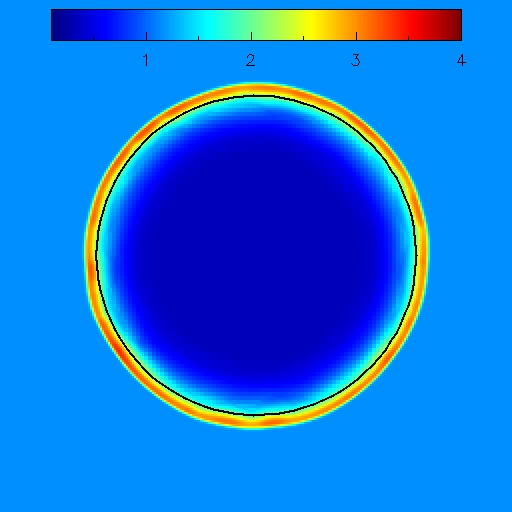}
\centering \includegraphics[width=0.33\textwidth]{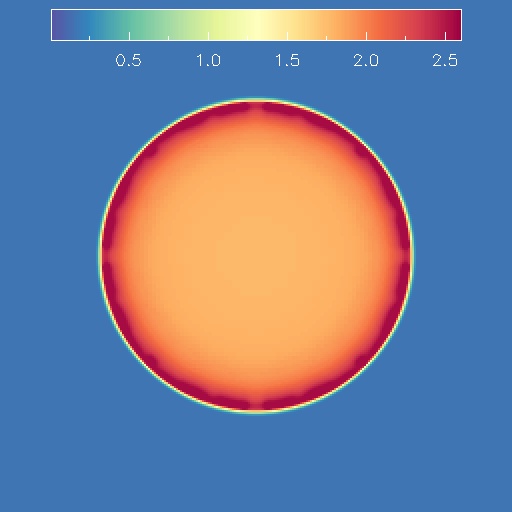}
\centering \includegraphics[width=0.33\textwidth]{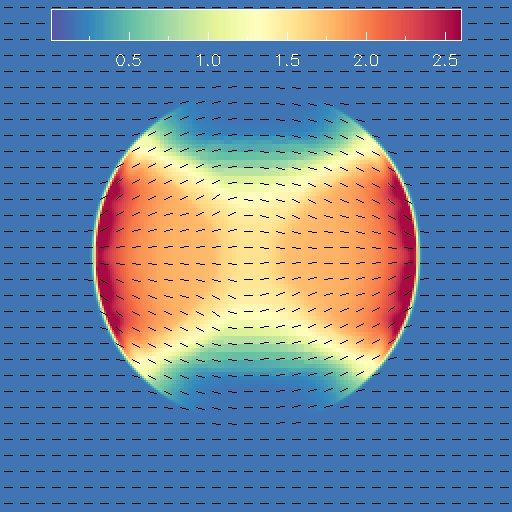}
\centering \includegraphics[width=0.33\textwidth]{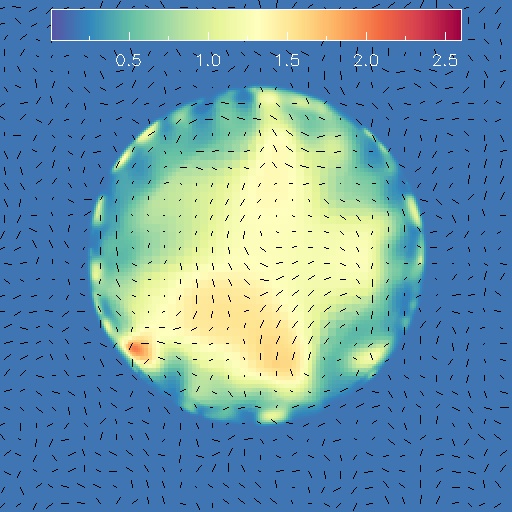}
\caption{Sedov explosion with accelerated CRs with $\eta=0.5$ (left panels), and obliquity-dependent acceleration efficiency $\eta=0.5\zeta(\theta_{\rm B})$ with either a uniform magnetic field (middle panels) or a random magnetic field (right panels). The top and bottom panels show respectively slices of density and CR pressure at time $t=0.05$, with the solid circle line indicating the position of the Sedov shock front for the exact solution with $\gamma_{\rm e}=7/5$, which are reproduced in all panels to guide the eye throughout (the random magnetic field configuration is better fitted with $\gamma_{\rm e}=1.55$), and with magnetic unit vectors overplotted as black segments (the length scale of the random magnetic field corresponds to the size of two large arrows). 
In the simulation without obliquity dependent acceleration, CR production is close to uniform in the shell except for small numerical grid artefacts. 
With obliquity dependency, CRs accumulate in polar caps for a uniform magnetic field, and in small patches for the random magnetic field corresponding to the length scale of the field. 
The position and  shape of the shell are also affected by the presence and the configuration of the magnetic field with respect to the obliquity-independent case.
}
\label{fig:sedov_image}
\end{figure*}

\subsubsection{Cosmic-ray acceleration with constant efficiency}

In this test we set up the previous 1D Sod shock tube test with Mach number  $\mathcal M=10$,  and allowed for CR acceleration with a constant efficiency of $\eta=0.5$ (the exact Mach number accounting for CRs added at the shock is $\mathcal M=9.56$ for this particular efficiency).
We used an adiabatic index for the thermal and CR components of respectively $\gamma=5/3$ and $\gamma_{\rm CR}=4/3$.
 All the Sod experiments were run without streaming and without radiative thermal or CR losses.
The analytical solution with accelerated CRs was provided by~\citealp{pfrommeretal17} (see their Appendix B).

Figure~\ref{fig:sod} shows the result of the numerical calculation where the analytical solution is nicely reproduced with the correct Mach number of $\mathcal{M}\simeq 9.56$ positioned at the shock front in one of the cell sampling the numerically broadened discontinuity.
Right after the shock discontinuity, in the post-shock region, the thermal pressure shows a few cells that overshoot the expected value.
This effect is due to our choice of depositing the accelerated CR energy density a few cells beyond the exact shock location (a strategy we employ to avoid  the $P{\rm d}V$ compression).
Apart from this expected effect, pressures, velocity, density, and the effective adiabatic index of the gas are accurately reproduced.

\subsubsection{Cosmic-ray acceleration with magnetic obliquity dependency}

In this Sod test, we let the acceleration efficiency $\eta(\theta_{\rm B})$ vary with the pre-shock magnetic obliquity angle $\theta_{\rm B}$ and imposed $\eta=0.5\zeta(\theta_{\rm B})$ (the previous Sod test was run with $\theta_{\rm B}=0^\circ$, i.e. the efficiency was $\eta=\eta_0=0.5$).
We ran three experiments with $\theta_{\rm B}=30,45$, and $60^\circ$ (i.e. $\zeta\simeq0.95$, 0.5, and  0.05 respectively), starting with an initial magnetic field with components $(B_{x},B_{y},B_{z})=$ $(10^{-10},0,5.77\times 10^{-11})$, $(10^{-10},0,10^{-10})$, $(5.77\times10^{-11},0,10^{-10}),$ respectively.
Magnetic field magnitudes were chosen to be arbitrarily small so that the magnetic field had no dynamical impact on the gas (i.e. $B^2 \ll P$).
The results are shown in Fig.~\ref{fig:sod_theta}, where we see that the expected values of the CR pressure in the shock are  reproduced well for any of the adopted magnetic obliquity.
We note that the exact location of the shock jump is modified, due to the modified shock velocity, which is governed by the effective adiabatic index in the shock that depends on the amount of accelerated CRs.

\begin{figure}
\centering \includegraphics[width=0.45\textwidth]{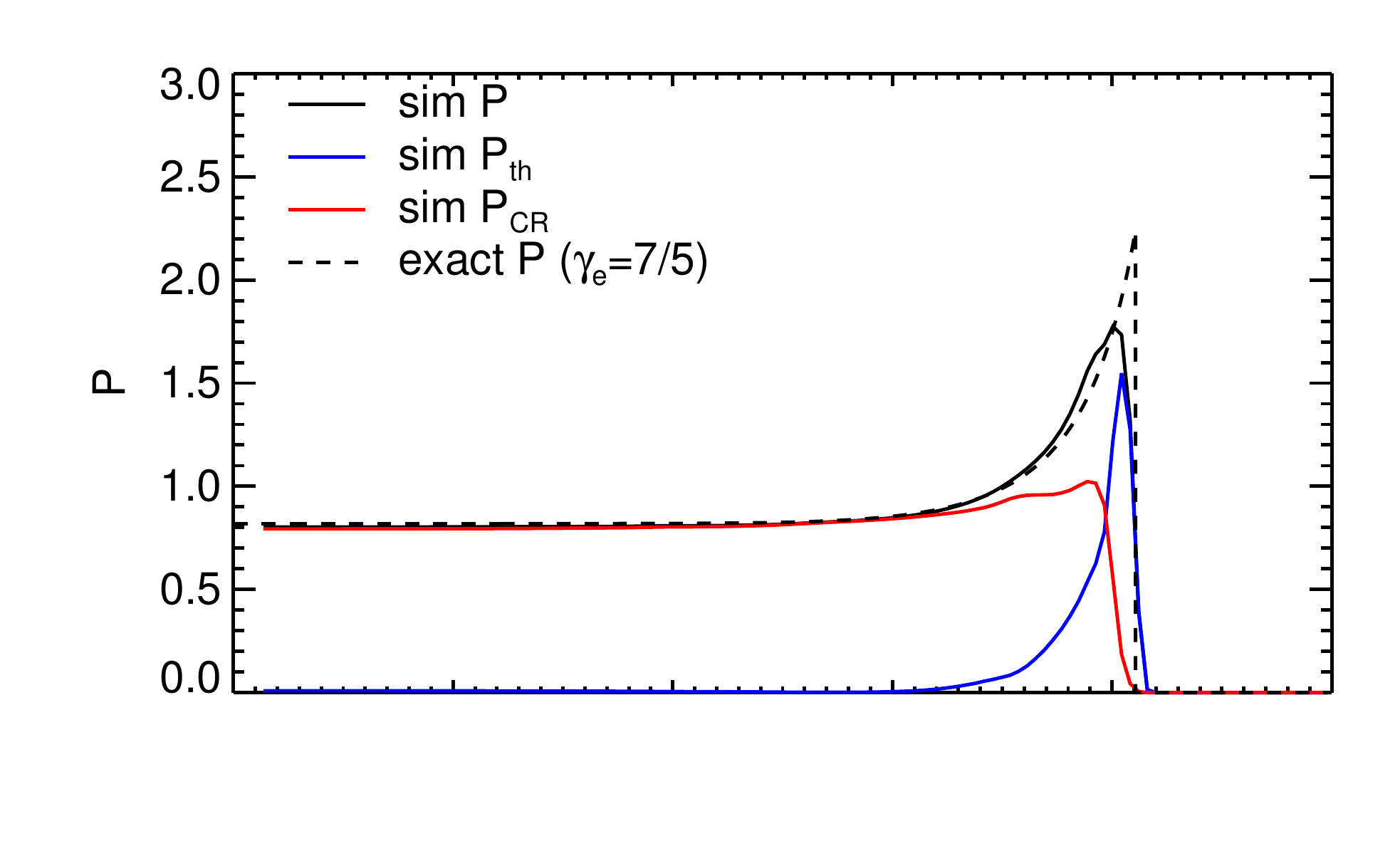}\vspace{-1.3cm}
\centering \includegraphics[width=0.45\textwidth]{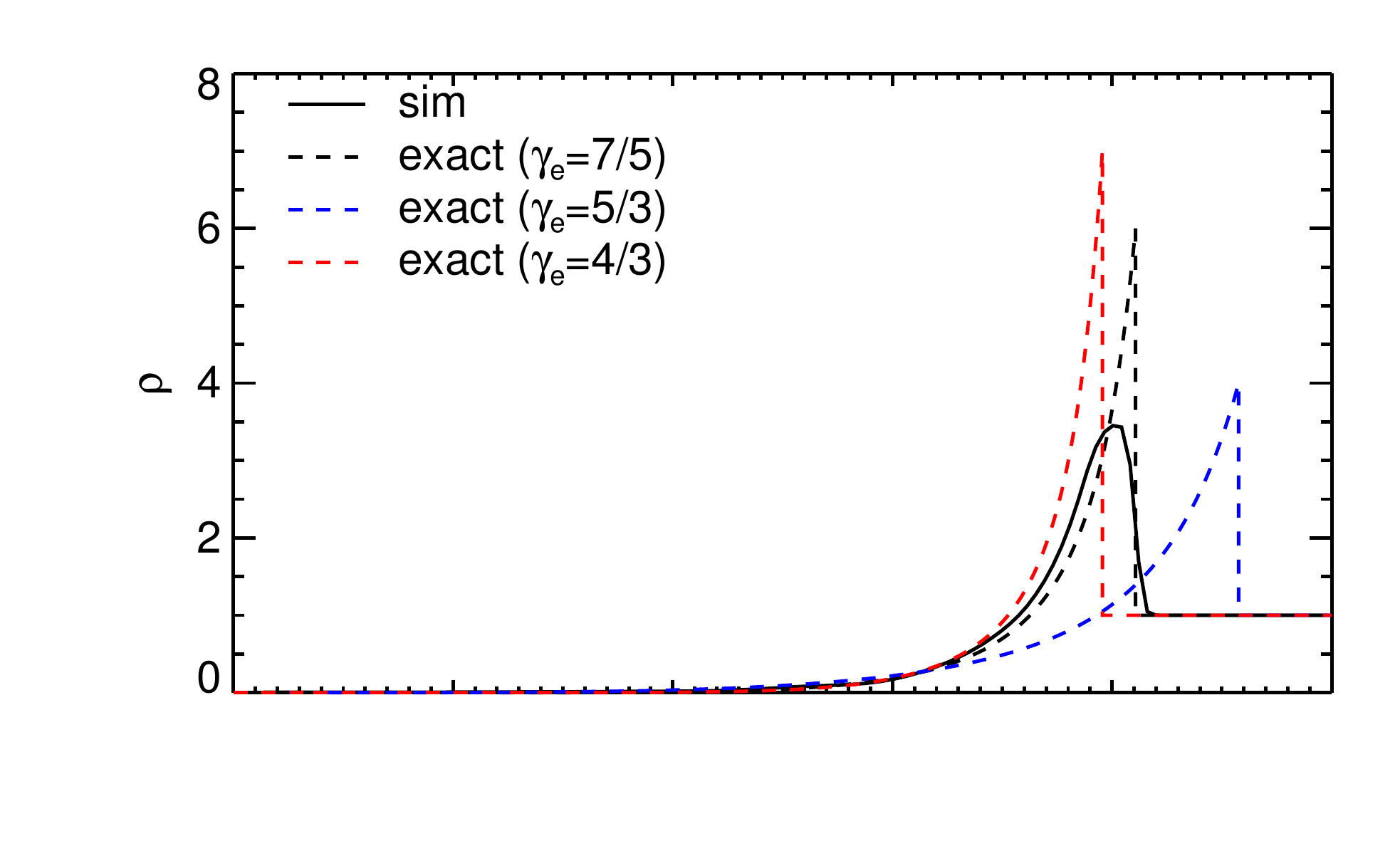}\vspace{-1.3cm}
\centering \includegraphics[width=0.45\textwidth]{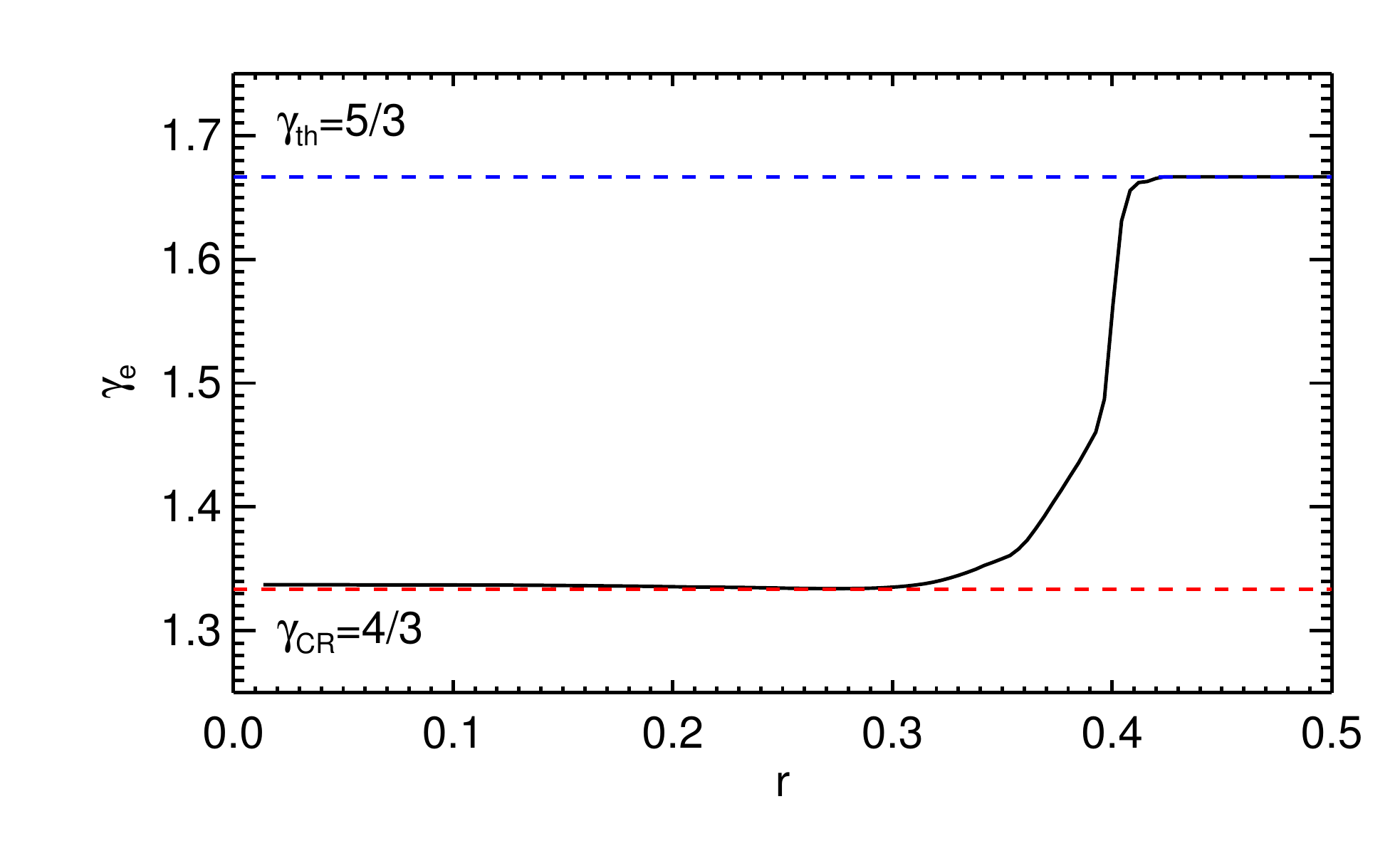}
\caption{Spherically averaged radial profiles for the 3D Sedov explosion with CR acceleration with constant acceleration efficiency of $\eta=0.5$ of the pressure (blue: thermal pressure, red: CR pressure), density, and effective adiabatic index of the thermal--CR mixture from top to bottom at time $t=0.1$. Solid lines stand for the result of the numerical simulation, while the dashed lines of  the pressure and density plots are the exact solution of the self-similar profile for an effective adiabatic index of $7/5$ in black (the exact density profile for $\gamma=5/3$ is also shown as a dashed blue line). The blue and red dashed lines in $\gamma_{\rm e}$ stand for the adiabatic index used for the thermal and CR component, respectively. The thermal--CR mixture produces an explosion similar to a Sedov solution with effective adiabatic index of $\gamma_{\rm e}=7/5$, which delays the position of the shock due to the lower pressure work exerted by the shocked shell.}
\label{fig:sedov_profiles}
\end{figure}

\begin{figure}
\centering \includegraphics[width=0.45\textwidth]{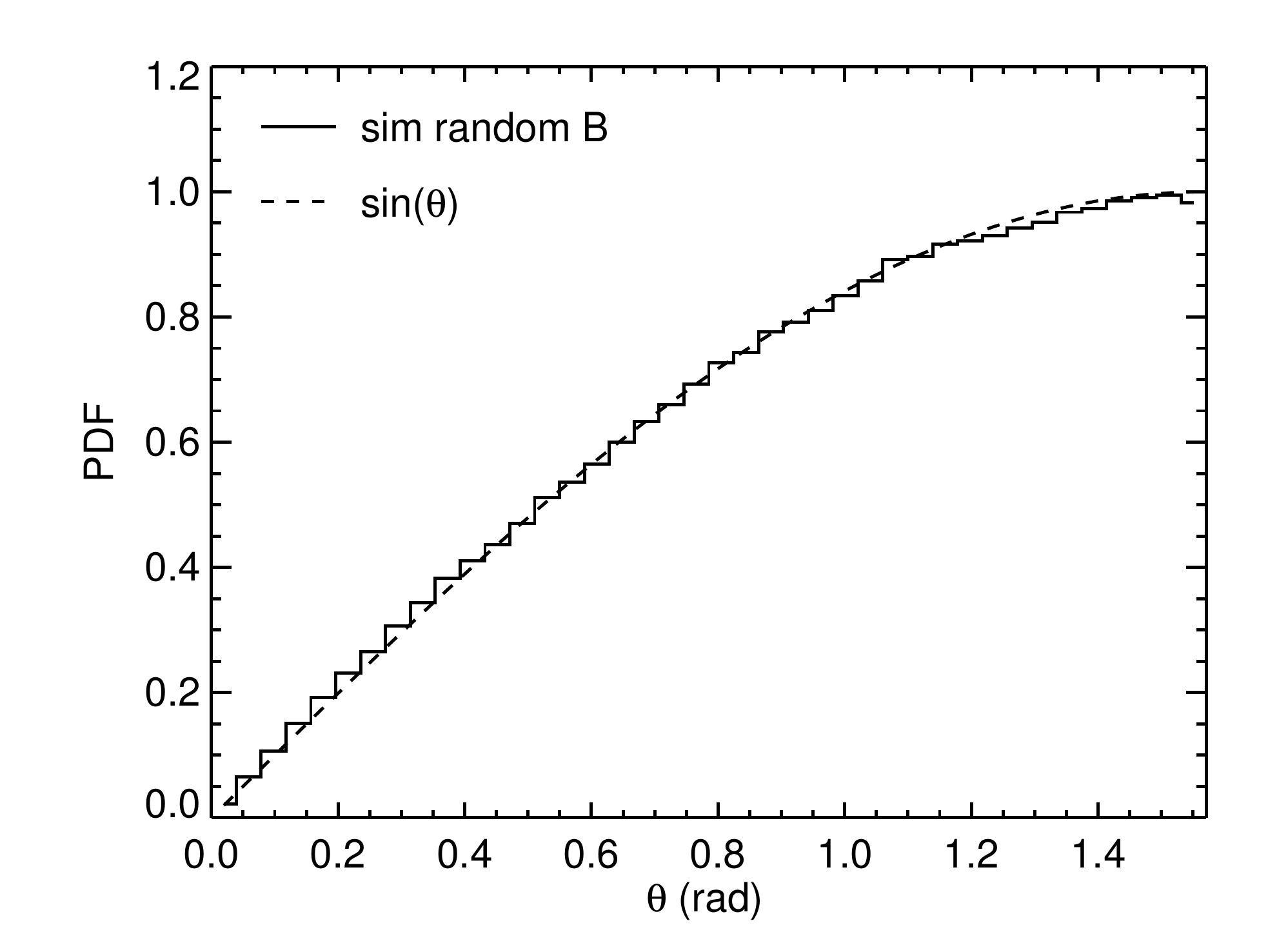}
\caption{Stacked PDF of the magnetic obliquity in the Sedov experiment between $t=0.05-0.1$ for the random magnetic field configuration (solid histogram), compared to the random distribution (black dashed line). The distribution of magnetic obliquity is compatible with a purely random field as expected, thus  leading to a reduced efficiency of $<\zeta>=0.302$.}
\label{fig:thetapdf_sedov}
\end{figure}

\begin{figure}
\centering \includegraphics[width=0.45\textwidth]{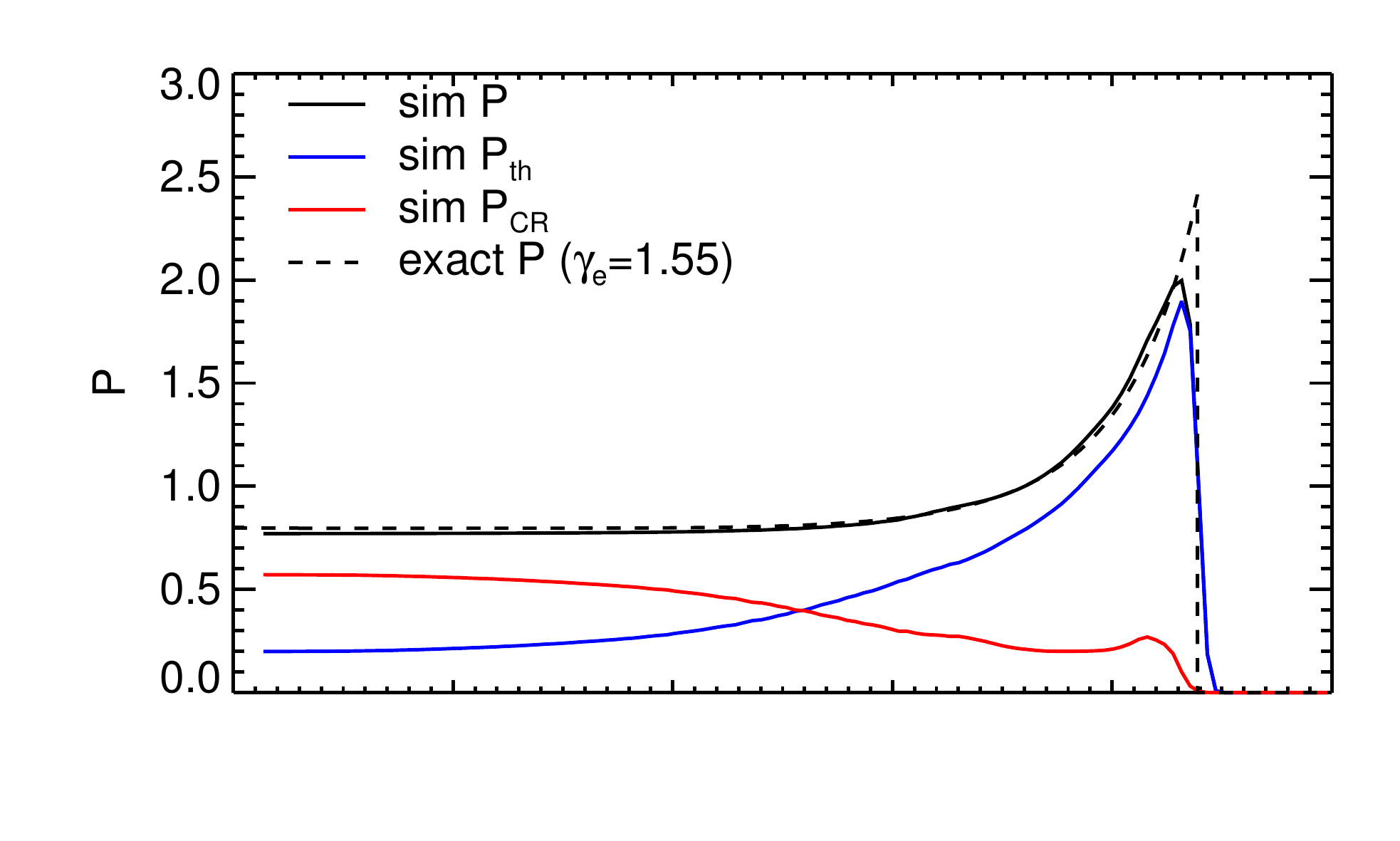}\vspace{-1.3cm}
\centering \includegraphics[width=0.45\textwidth]{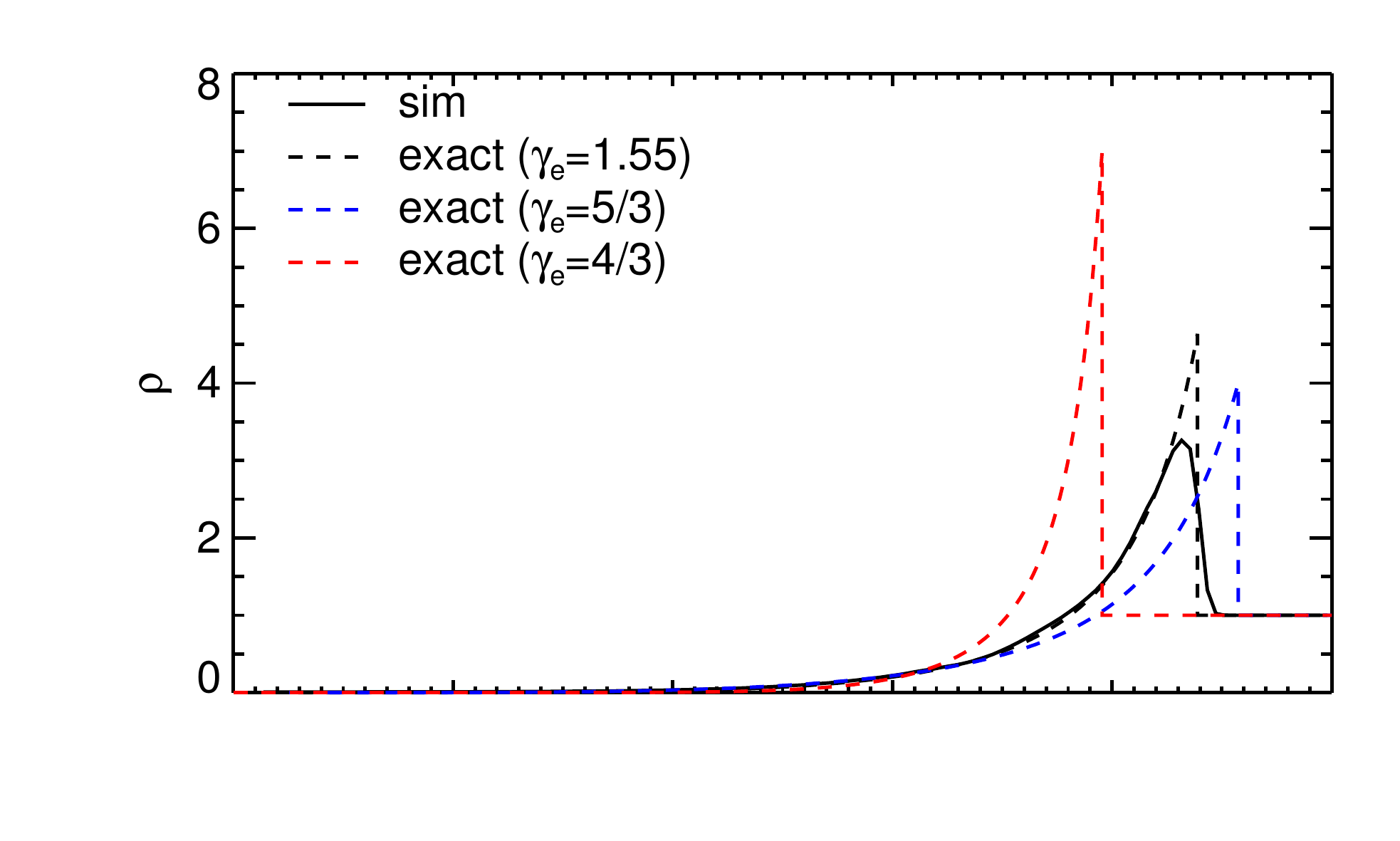}\vspace{-1.3cm}
\centering \includegraphics[width=0.45\textwidth]{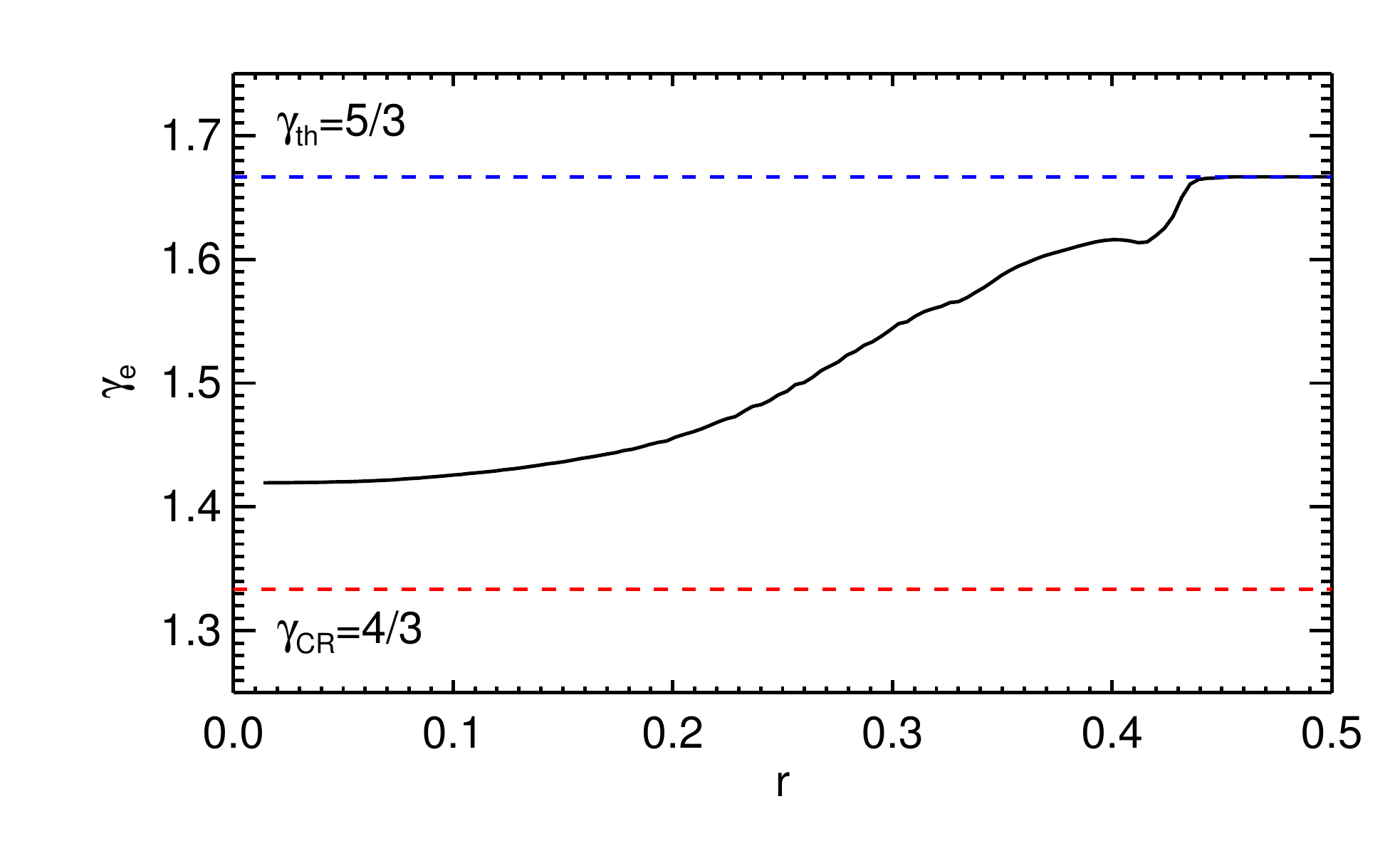}
\caption{Similar to Fig.~\ref{fig:sedov_profiles}, but  for the random magnetic field configuration and with obliquity dependency of the CR acceleration efficiency $\eta=0.5\zeta(\theta_{\rm B})$. Here the Sedov profile is better fitted with an effective adiabatic index of $\gamma_{\rm e}=1.55$.}
\label{fig:sedov_profiles_brand}
\end{figure}

\subsection{Three-dimensional Sedov explosion}
\label{section:sedov}

We set up a 3D Sedov explosion with the following unitless values: a background at rest with gas density of $\rho=1$, $P_{\rm th}=10^{-4}$, and a point-like explosion of energy $E_{\rm th}=1$ spread over the eight central cells in a box of size unity\footnote{These adopted unitless values can correspond to e.g. a SN explosion of $1.1\times10^{\rm 51}\,\rm erg$ in a background medium of density $n=1\, \rm H\, cm^{-3}$, sound speed $c_\mathrm{s}=0.6\, \rm km\, s^{-1}$, and a box length of 45 pc.}.
There are no CRs initially, and only those accelerated into the shock with a constant acceleration efficiency of $\eta=0.5$ will necessarily contribute to the CR distribution.
The adiabatic index of the thermal component is $\gamma=5/3$, and  $\gamma_{\rm CR}=4/3$ for CRs.
In a box of size unity, we start with a base grid of level 6 and allow for 2 extra levels of refinement wherever the cell-to-cell density and pressure variations are larger than 20 \%\ and 50 \%, respectively.
The criterion for density is used only where the gas density is higher than that of the background in order to avoid excessive refinement into the hot interior, and instead we  focus on the shocked swept-up shell material.
For this particular test it is customary to employ a more diffusive solver than HLLD (or Harten--Lax--van Leer--Contact for a pure hydro run) to avoid the formation of the carbuncle phenomenon in shocked cells around the x-, y-, or z-axis of the box, hence, we use, here, the Lax-Friedrich approximate Riemann solver.
All Sedov experiments are run without streaming and without radiative thermal or CR losses.

Figure~\ref{fig:sedov_image} (left panels) shows the density and CR pressure in a thin slice through the centre of the explosion at time $t=0.05$.
The swept-up material accumulates in a thin shocked layer of gas where CRs are accelerated and they propagate backward through a reverse shock in the bubble interior.
We can see finger-like features in the shocked material, which are produced by the discretised nature of the grid;  amongst the post-shock cells receiving the accelerated CR energy, some of them can indeed receive energy from several shock cells, while some others receive it only once. 
We note that~\cite{pfrommeretal17} also noticed this effect in their unstructured mesh code, the difference is that their features are randomly located in angle, while here, due to the structured cartesian nature of our grid, these features follow some $\pi/2$ periodic pattern.

As expected, due to the high adopted value of acceleration efficiency $\eta=0.5$, there is a very significant amount of CRs produced into the dissipation layer of the shock as seen in the spherically averaged radial profiles from Fig.~\ref{fig:sedov_profiles}.
The pressure in the shock layer is a mixture of CRs and thermal particles, while the CR pressure completely dominates the total pressure in the diffuse bubble interior.

It leads to a sharp transition of the effective adiabatic index of the gas from purely thermal outside of the explosion $\gamma_{\rm e}=\gamma$ to purely CR-like in the diffuse bubble $\gamma_{\rm e}=\gamma_{\rm CR}$.
What matters for the shock dynamics is the effective adiabatic index in the swept-up shock layer that can be inferred from the exact Sedov shock dynamics given a value of $\gamma_{\rm e}$.
For analytical guidance, with enthalpy arguments~\cite{chevalier83} provides the solution for the effective adiabatic index as a function of the fraction of CR pressure $w=P_{\rm CR}/P_{\rm tot}$ in  the shocked shell (not to be confused with the acceleration efficiency) 
\begin{equation}
\gamma_{\rm e}=\frac{5+3w}{3(1+w)} \, 
\end{equation}
for $\gamma_{\rm CR}=4/3$. 
In agreement with~\cite{pfrommeretal17}, we find that for the same set-up, an effective adiabatic index in the shock of $\gamma_{\rm e}=7/5$ for the exact solution leads to a good recovery of the numerical solution in both total pressure and density, though the maximum values are less pronounced at the shock because of the limited resolution. 
Increasing the resolution naturally captures  the shock profile more faithfully.

We ran two extra simulations with the acceleration efficiency depending on magnetic obliquity $\eta=0.5\zeta(\theta_{\rm B})$ and changing from an initial initially uniform magnetic field with $(B_x,B_y,B_z)=(10^{-10},0,0)$ or a random magnetic field configuration (see Appendix~\ref{app:brandom} for details) with a typical coherence length of $\lambda_{\rm B}=1/16$ and a similar magnitude of $10^{-10}$.
For the uniform magnetic field configuration, CRs are accelerated around polar caps along the x-axis of the box with maximum efficiency, and go to zero along the y-axis (or z-axis) as a result of magnetic obliquity (see middle panels of Fig.~\ref{fig:sedov_image}).
 It results in  an ellipsoid shape of the explosion: the position of the shell where CR acceleration is close to zero (y- and z-axes) is further away than where CRs are produced (x-axis) as a result of the higher (resp. lower) effective adiabatic index of the gas mixture in the shell.
We note that the exact shape of the ellipsoid is a function of the obliquity-independent part of the acceleration efficiency:  the larger $\xi$ is, the more stretched  the explosion is~\citep[see][for a thorough analysis of this effect]{paisetal18}.
As expected, the density is also higher along the x-direction than along the y-direction (z-direction) as a result of the dependency of the density jump to the adiabatic index of the gas (for strong shocks, $\mathcal{R}_{\rho}=4$ for $\gamma_{\rm e}=5/3$ and $\mathcal{R}_{\rho}=6$ for $\gamma_{\rm e}=7/5$).

Finally, the random magnetic field set-up shows a shell mass distribution close to spherical with significant fluctuations with angle (right panels of Fig.~\ref{fig:sedov_image}).
It reflects the underlying patchy acceleration and distribution of CR pressure in the swept-up shock layer.
On average, the acceleration efficiency is reduced by a factor $<\zeta>=\int_0^{\pi/2} \zeta(\theta_{\rm B}) \sin \theta_{\rm B} d\theta_{\rm B}\simeq0.302$ for a purely random upstream magnetic field orientation (see Fig.~\ref{fig:thetapdf_sedov}) compared to the simulation without obliquity dependency, and thus to an effective acceleration parameter of $\eta_{\rm e}\simeq 0.15$. Therefore, there is a smaller amount of CRs produced in the shock, and as expected from~\cite{chevalier83}~\citep[see also][]{castroetal11, bell15}, the exact solution is now better reproduced for a lower effective adiabatic index of $\gamma_{\rm e}=1.55$ (see Fig.~\ref{fig:sedov_profiles_brand}) and leads to a shock front in advance compared to the obliquity-independent simulation.

\section{Turbulent box of the interstellar medium}
\label{section:ism}

\begin{figure}
\centering \includegraphics[width=0.45\textwidth]{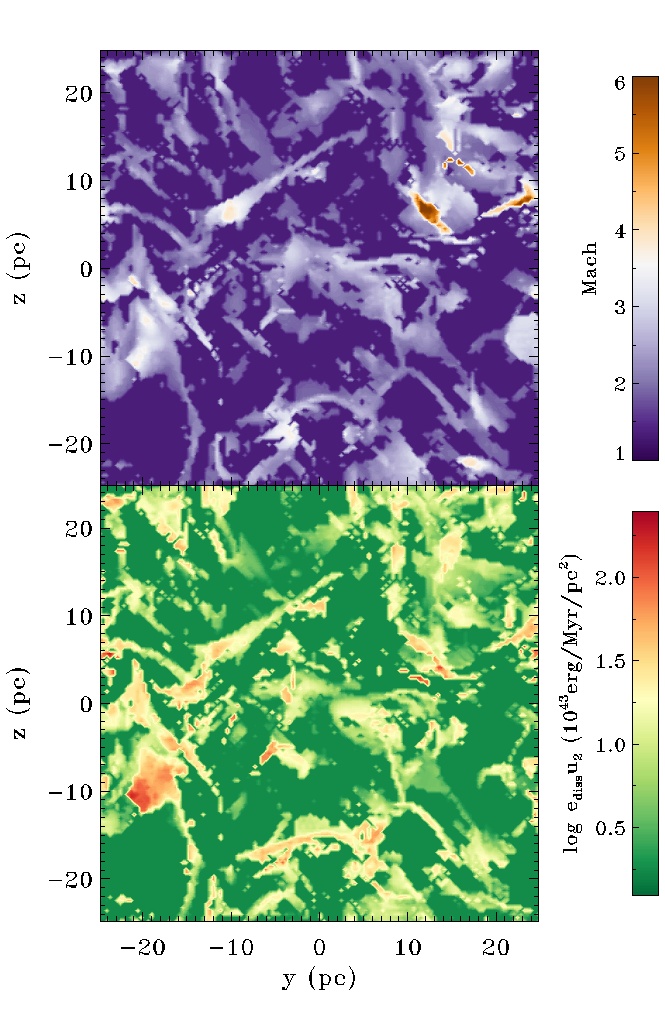}
\caption{Projection of the Mach number $\mathcal{M}$ (top) and dissipated energy flux $e_{\rm diss}u_2$ (bottom) for the Streaming turbulent box, with $\eta_0=0.1$ and $\xi(\mathcal{M},X_{\rm CR})=1$, at time $t=10 \,\rm Myr$ over a box thickness of half the size of the box centred on the middle of the box. Shocks are driven in sheets with a bulk of the Mach number of moderate values $\mathcal{M}\simeq3$--$4$.}
\label{fig:machediss_ism}
\end{figure}

\begin{figure*}
\centering \includegraphics[width=\textwidth]{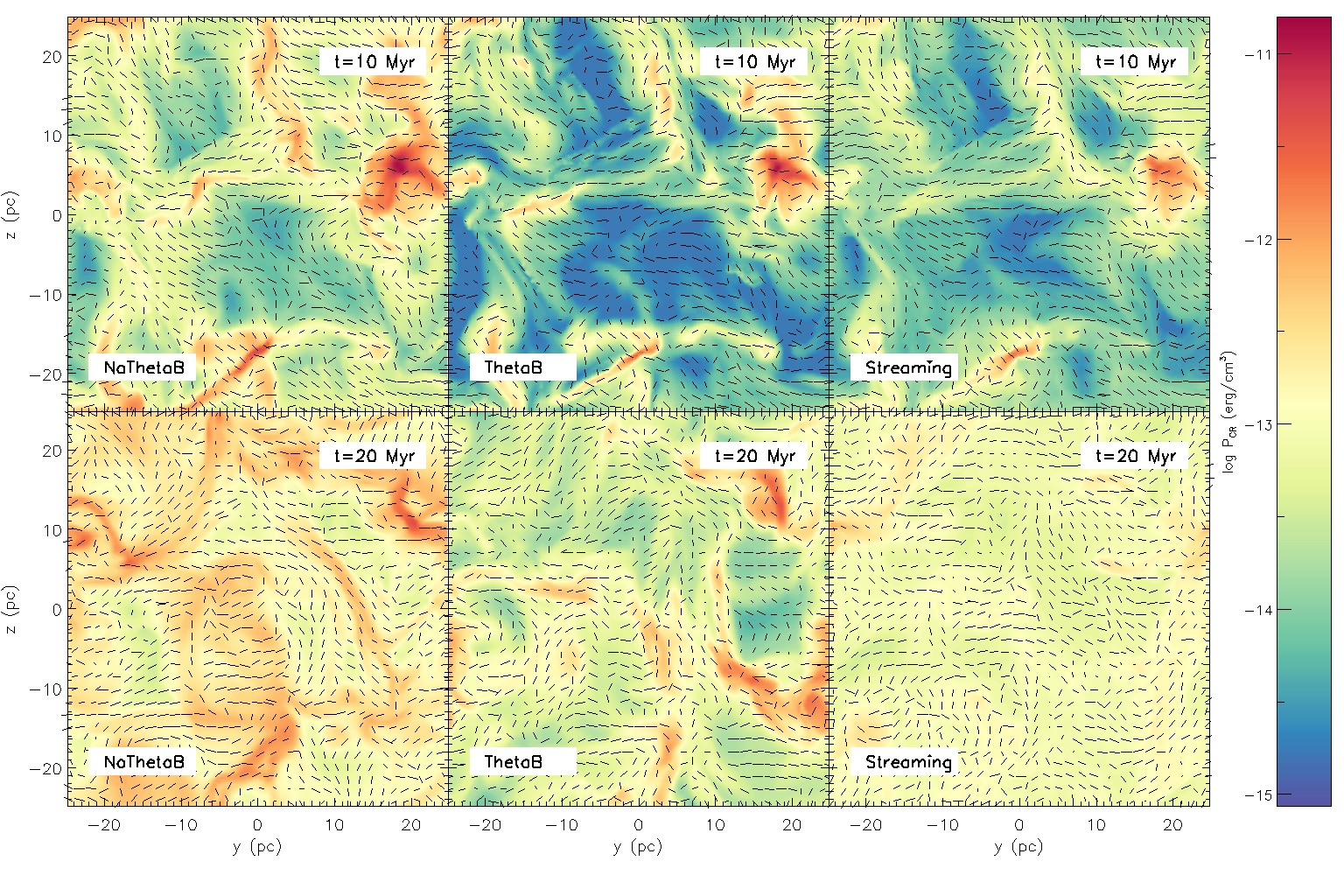}
\caption{Cosmic-ray  pressure maps of the turbulent box simulation, with $\eta_0=0.1$ and $\xi(\mathcal{M},X_{\rm CR})=1$, in a thin plane within the x-plane of the middle of the box at time $t=10\, \rm Myr$ (top panels) and $t=20 \, \rm Myr$ (bottom panels) for the simulation without CR streaming and without (left panels) or with (right panels) obliquity dependency for CR acceleration, and with obliquity and CR streaming (right panels). The black segments depict the orientation of the unitary magnetic vectors. The simulation without obliquity builds the CR pressure faster. The presence of the streaming instability allows for a more uniform distribution of CRs in the simulated volume.}
\label{fig:nice_ism}
\end{figure*}

\begin{figure}
\centering \includegraphics[width=0.45\textwidth]{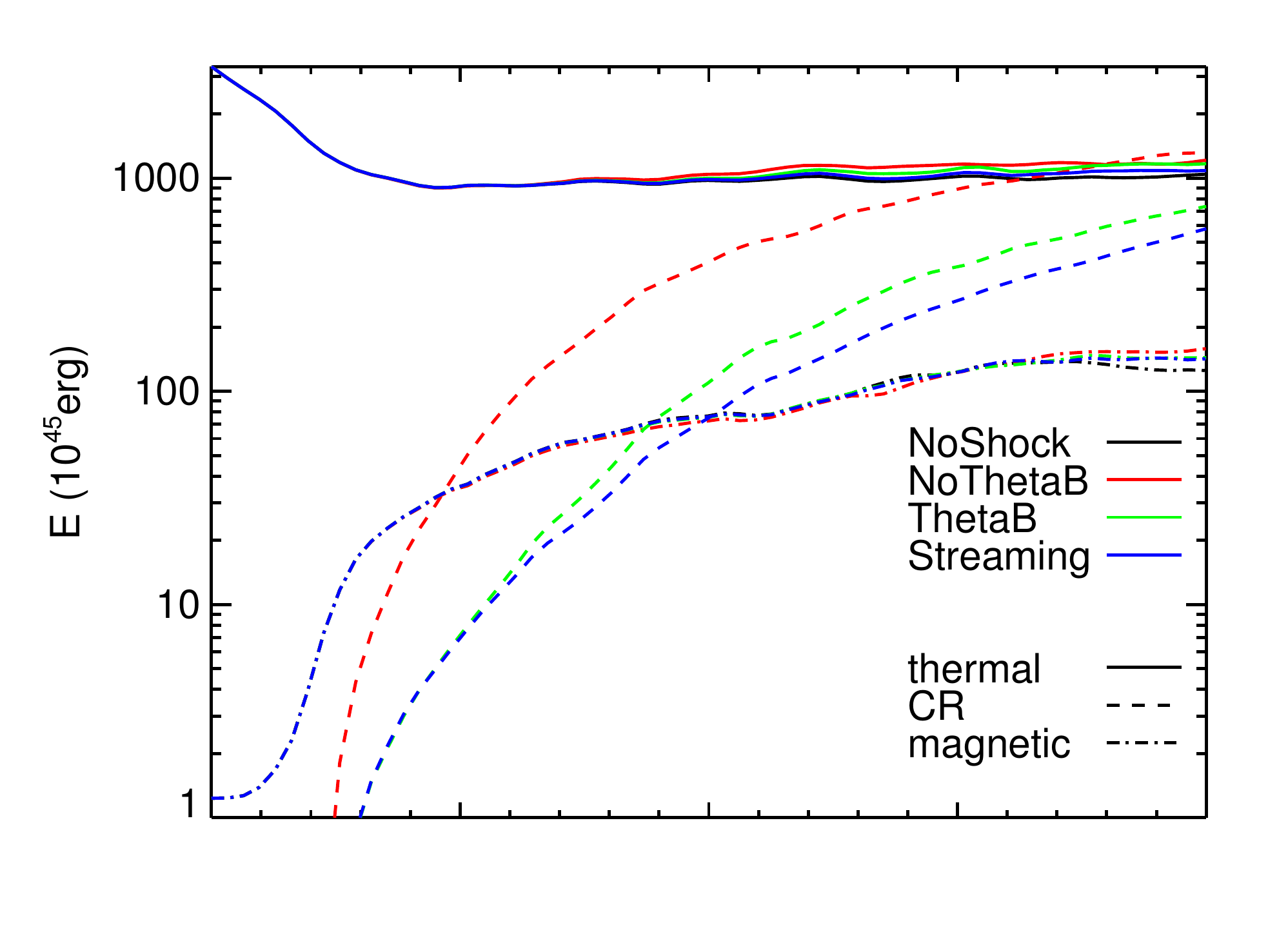}\vspace{-1.25cm}
\centering \includegraphics[width=0.45\textwidth]{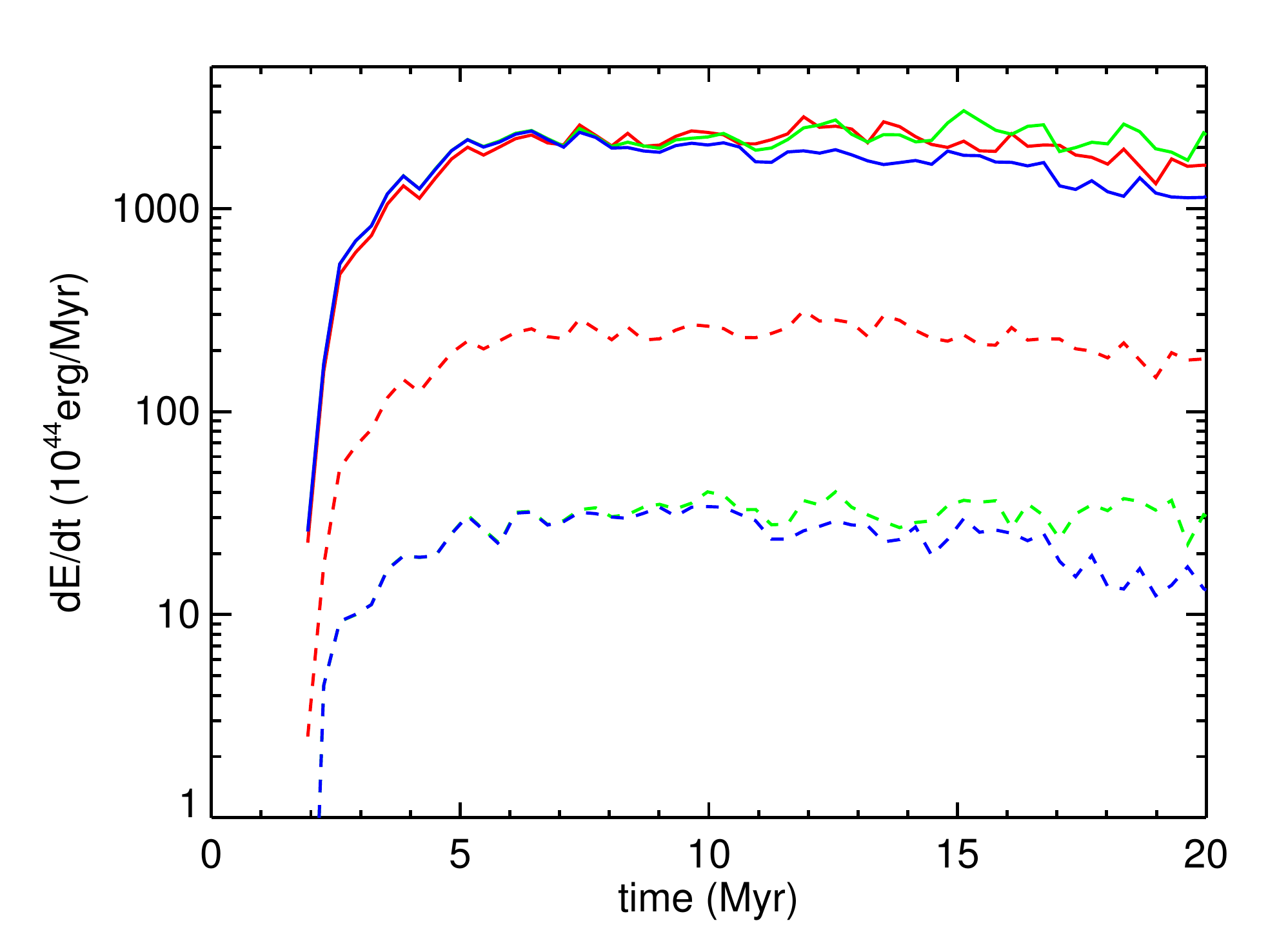}
\caption{Top panel: Time evolution of total thermal (solid lines), CR energy (dashed lines), and magnetic energy (dot-dashed lines) in the simulated turbulent ISM boxes for the simulations without shock-acceleration (black), without CR streaming, and without (red) or with (green) obliquity dependency for CR acceleration, and with obliquity and CR streaming (blue) with $\eta_0=0.1$ and $\xi(\mathcal{M},X_{\rm CR})=1$. Bottom panel: Evolution of the dissipated thermal (solid) and CR (dashed) energy rates at shocks for the same simulations. }
\label{fig:eevol_ism}
\end{figure}

\begin{figure}
\centering \includegraphics[width=0.45\textwidth]{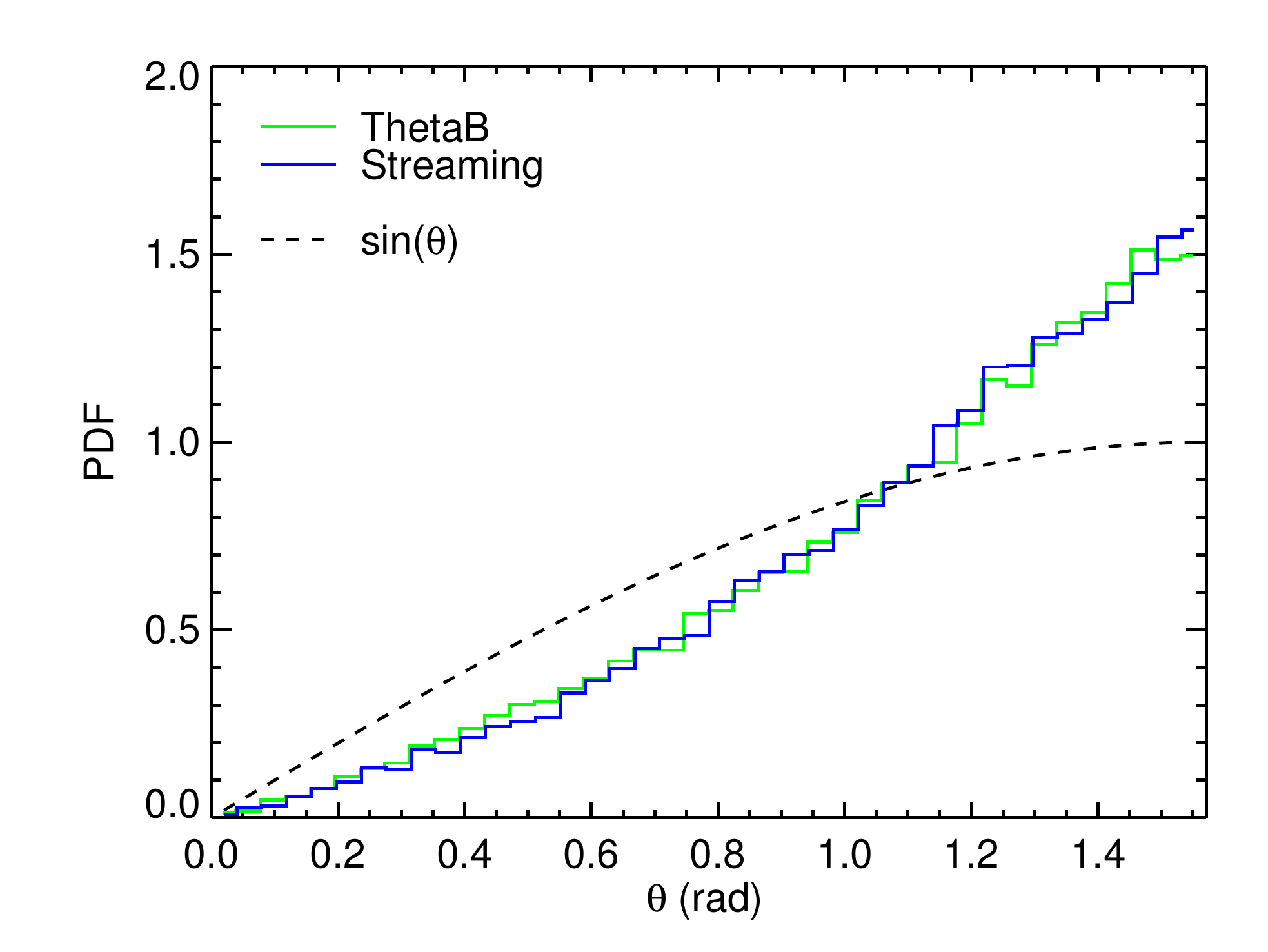}
\caption{Probability density function of the magnetic obliquity in the ISM boxes, and with $\eta_0=0.1$ and $\xi(\mathcal{M},X_{\rm CR})=1$, at time $t=20 \, \rm Myr$ with CR streaming (blue) or without (green), compared to the random distribution in black dashed. Those simulations are more likely to have magnetic field perpendicular to the normal of shocks than for a random distribution, therefore, lowering the CR acceleration efficiency compared to the averaged random distribution, i.e. $<\zeta>=0.302$.}
\label{fig:thetapdf_ism}
\end{figure}

\begin{figure}
\centering \includegraphics[width=0.45\textwidth]{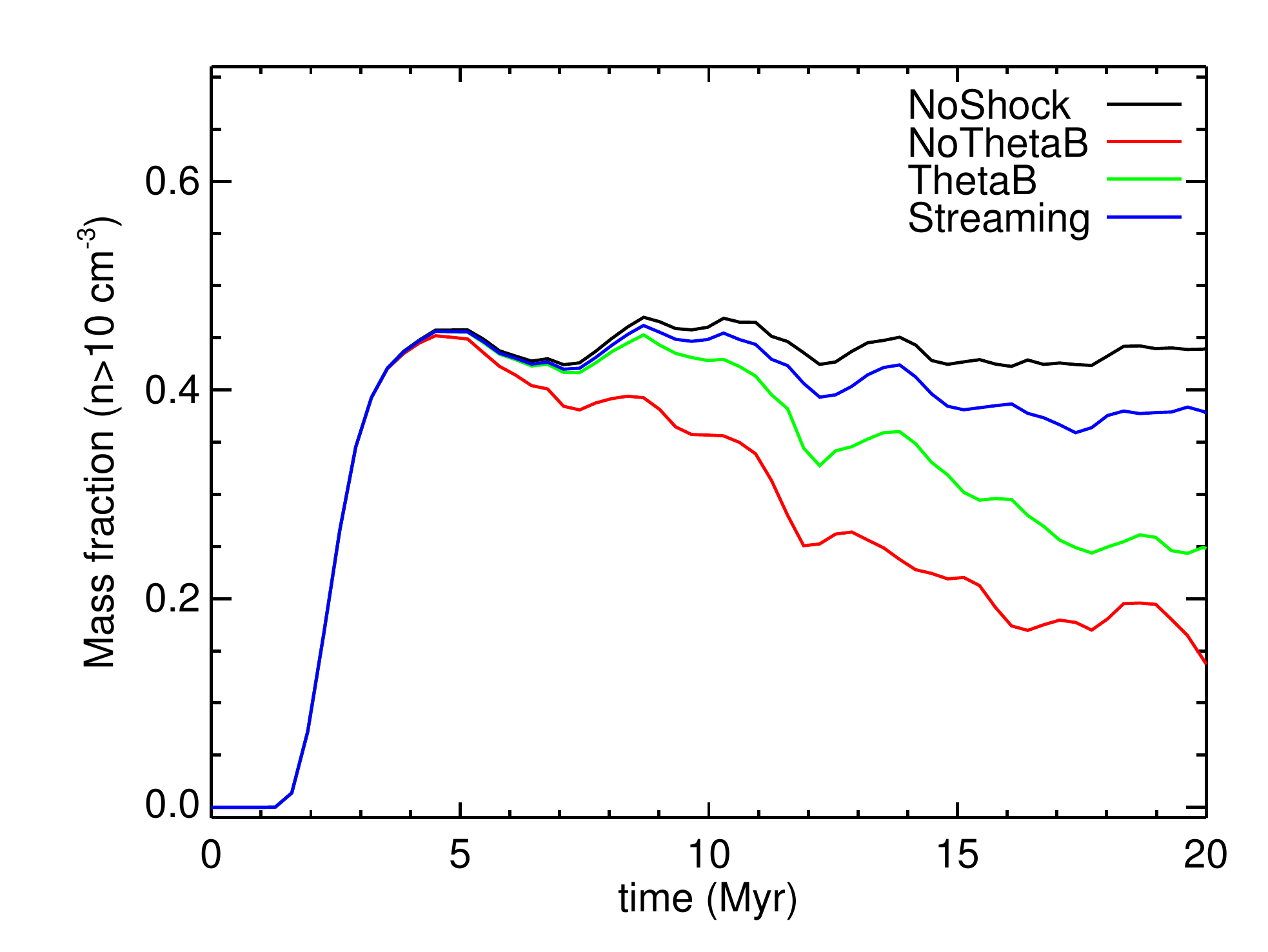}
\caption{Time evolution of the mass fraction of dense gas in the simulated turbulent ISM boxes for the simulations without shock-acceleration (black), without CR streaming, and without (red) or with (green) obliquity dependency for CR acceleration, and with obliquity and CR streaming (blue), and with $\eta_0=0.1$ and $\xi(\mathcal{M},X_{\rm CR})=1$. }
\label{fig:nevol_ism}
\end{figure}

We ran turbulent interstellar medium (ISM) boxes in the same spirit of~\cite{commerconetal19} except that here we started with negligible CR pressure ($10^{-10}$ that of the thermal pressure) and let it build through the turbulence-generated shocks.
The simulations have a uniform $128^3$ cartesian resolution in a box of 50 pc, leading to a spatial resolution of $0.4 \, \rm pc$.
The initial gas density and temperature are $2\,\rm cm^{-3}$ and $4460\,\rm K,$ respectively, with a mean molecular weight of $\mu=1.4$ assumed throughout.
 We started with an initial thermal pressure of $P_{\rm th,0}=1.2\times 10^{-12}\,\rm erg\, cm^{-3}$.
The initial magnetic field was uniform and was set up in the x-direction of the box with a magnitude of $0.1\,\rm \mu G$, leading to a plasma beta parameter of $\beta=P_{\rm th,0}/P_{\rm mag,0}\simeq3\times 10^3$.
We did not allow for self-gravity of the gas or for any refinement. 
Cooling proceeded on the thermal component following~\cite{audit&hennebelle05}, while we neglected the role of Coulomb and hadronic losses of CR protons~\citep{ensslinetal07,guooh08}.

The turbulence is forced at all times with an injection scale of $k_{\rm turb}=2$ (i.e. corresponding to half the size of the box) and with a parabolic  shape in the Fourier space $\tilde f(k)\propto 1-(k-k_{\rm turb})^2$ with $k$ sampled in the range $k=[1,3]$. The turbulence is applied intermittently with an auto-correlation time of $0.5\, \rm Myr$ and with a compression-to-solenoidal ratio of 1~\citep[see][for more details]{commerconetal19}.

\subsection{$\mathcal{M}$- and $X_{\rm CR}$-independent acceleration efficiency}

We start with a batch of simulations where the acceleration efficiency does not depend on $\mathcal{M}$ and $X_{\rm CR}$ (i.e. $\xi=1$).
We set up three different simulations: i) without CR acceleration (i.e. $\eta_0=0$, NoShock); ii) with CR acceleration and $\eta=\eta_0=0.1$ (i.e. where CR acceleration does not depend on magnetic, NoThetaB); iii) with CR acceleration and $\eta_0=0.1$ (i.e. where CR acceleration depends on magnetic obliquity, ThetaB); and with $\eta_0=0.1$ and CR streaming (Streaming).
We note that we use rather large values of CR acceleration efficiencies given the moderate Mach numbers of only 2-4~\citep[e.g.][]{kang&jones05, kang&ryu13} obtained in that experiment. 
This somewhat reflects the more typical SN-generated CR acceleration efficiencies corresponding to much larger values of the shock Mach number than we can capture here with this simplified  set-up.
For the sake of a testable set-up for our new implemented algorithm, these values allow us to reach an appreciable amount of CR energy density in the simulated volume over a few turbulent crossing times $t_{\rm cross}=6.7\,\rm Myr$, where it is the box length divided by the rms velocity $u_{\rm rms}=7.3\,\rm  km\, s^{-1}$ (here measured at $t=20 \,\rm Myr$ for the Streaming run).

Shocks are driven in sheets with moderate Mach numbers of $\mathcal{M}\simeq3$--$4$, as can be seen in Fig.~\ref{fig:machediss_ism} for the Streaming run (other simulations show similar features) at time $t=10\, \rm Myr$, which dissipates the energy of shocks with a typical range of flux values of $e_{\rm diss}u_2\simeq10^{44}$--$10^{45}\,\rm erg\,Myr^{-1}\,pc^{-2}$.
Figure~\ref{fig:nice_ism} shows maps of the CR pressure at two different times $t=10$ and $20\, \rm Myr$ for the simulation NoThetaB, ThetaB, and Streaming.
At $t=10\,\rm Myr$ the CR pressure has already built up to appreciable levels thanks to turbulence-generated shocks in the box, with clustered regions of pressure at levels similar to or above the initial thermal pressure ($P_{\rm th,0}\simeq10^{-12}\,\rm erg\, cm^{-3}$).
The NoThetaB simulation has, as expected, the highest values of CR pressure since CR acceleration efficiency is always equal to $\eta=0.1$, while in the two other runs it can only reach this value for a perfectly aligned pre-shock magnetic field with the normal to the shock. 
At this early stage of the simulation, the effect of streaming is still very moderate on the CR pressure distribution. It reduces the range of the lowest and highest values of pressure mimicking the effect of a diffusion process; nonetheless, the geometrical features are easily recognizable between the ThetaB and Streaming runs (and NoThetaB as well).

Figure~\ref{fig:eevol_ism} (top panel) shows the thermal and CR energies in the simulated volumes as a function of time.
The total thermal energy in the box is quickly reduced in 3 Myr by nearly a factor of 3 with very negligible differences by the end of the simulation between the four simulations.
The total CR energy builds up almost linearly with time as a result of nearly constant dissipated energy and acceleration efficiency over time, once passed the first 5 Myr.
This CR pressure provides a support to the total pressure close to the thermal pressure, if not above (NoThetaB case at $t=20\,\rm Myr$).
The magnetic energy quickly increases early on and saturates at a plasma beta $\beta\simeq10$ similar for the four different simulations.
We note that this level of magnetic field is crucial for the CR streaming to have an appreciable effect on the CR pressure distribution as the streaming velocity scales with the Alfv\'en velocity.

As we discussed in section~\ref{section:sedov}, the average obliquity-dependent part of the CR acceleration efficiency must be $<\zeta>\simeq0.302$ for a purely random field, which seems supported by the apparent randomness of magnetic vectors (white arrows in Fig.~\ref{fig:nice_ism}), but we  show that this is not the case.
Figure~\ref{fig:eevol_ism} (bottom panel) shows the dissipated energy per unit time in the form of thermal or CR energy.
Dissipated thermal energies are very similar for the three simulations, although there is  a slight deviation at late times for the Streaming run.
However, the dissipated CR energy shows a larger than a factor 3 difference between the non-$\theta_{\rm B}$ and the $\theta_{\rm B}$ dependencies, closer to a factor 6-8 difference between the NoThetaB and ThetaB runs.
This is  indirect evidence that pre-shock magnetic fields are not randomly oriented, but show preferentially within-shock-plane orientations.
To clarify further, we measure the probability density function (PDF) of the obliquity for the ThetaB and Streaming runs at time $t=20\,\rm Myr$ in Fig.~\ref{fig:thetapdf_ism}, which shows that the PDF is skewed towards larger angles:  upstream magnetic fields are more likely to be perpendicular to the normal of shocks than for a random field, in agreement with the estimated reduced efficiency of CR acceleration.

We also note  that at time $t=10\,\rm Myr$, the CR energy density is a factor 2 lower with streaming, while the CR dissipated energy before $t\leq10\, \rm Myr$ is similar to that of the simulation without streaming.
Therefore, this difference in CR energy density is directly due to streaming (as opposed to streaming reducing shock strengths) putting CRs away from compressed regions (shocks or not) where the adiabatic compression can further enhance the overall CR pressure.

At time $t=20 \,\rm Myr$, the distributions of CR pressure (Fig.~\ref{fig:nice_ism}) in the three simulations differ very significantly.
While the NoThetaB and ThetaB runs look like a renormalised versions of one another, albeit with different specific locations of voids and plume-like features, the Streaming run has lost most of its CR structure  with a closer to uniform distribution of CR pressure in the box.

These distinct CR pressure evolutions and distributions lead to very important differences in the way the matter is compressed into overdense regions of the flow.
Figure~\ref{fig:nevol_ism} shows the time evolution of the mass fraction of dense gas, which is arbitrarily chosen at five times the initial gas density (i.e. for $n>10\,\rm cm^{-3}$, but the results are qualitatively independent of this choice).
Since only the thermal pressure is affected by radiative losses, which are larger at high gas densities, it is the CR pressure that accumulates in regions of high gas densities that can provide the support against compression.
Therefore, it shows that the simulations with the largest total CR energy are the simulations with the lowest amount of dense gas.
However, the streaming introduces a subtle but significant difference to this overall picture.
Since streaming smooths the CR pressure in the ISM, the high gas density is much less clustered for a given total energy in the box. 
At $t=20\,\rm Myr$ in the Streaming run, the total CR energy is indeed equal to that at $t=18\,\rm Myr$ in the ThetaB run; nonetheless, the mass fraction of dense gas is respectively 40 \%\ higher in the Streaming run.
Recast into an `effective' diffusion framework, we can deduce that streaming behaves like anisotropic diffusion with an effective diffusion coefficient to be determined through comparison with the corresponding simulations, which we defer to a  future work.

\subsection{$\mathcal{M}$- and $X_{\rm CR}$-dependent acceleration efficiency}

\begin{figure}
\centering \includegraphics[width=0.45\textwidth]{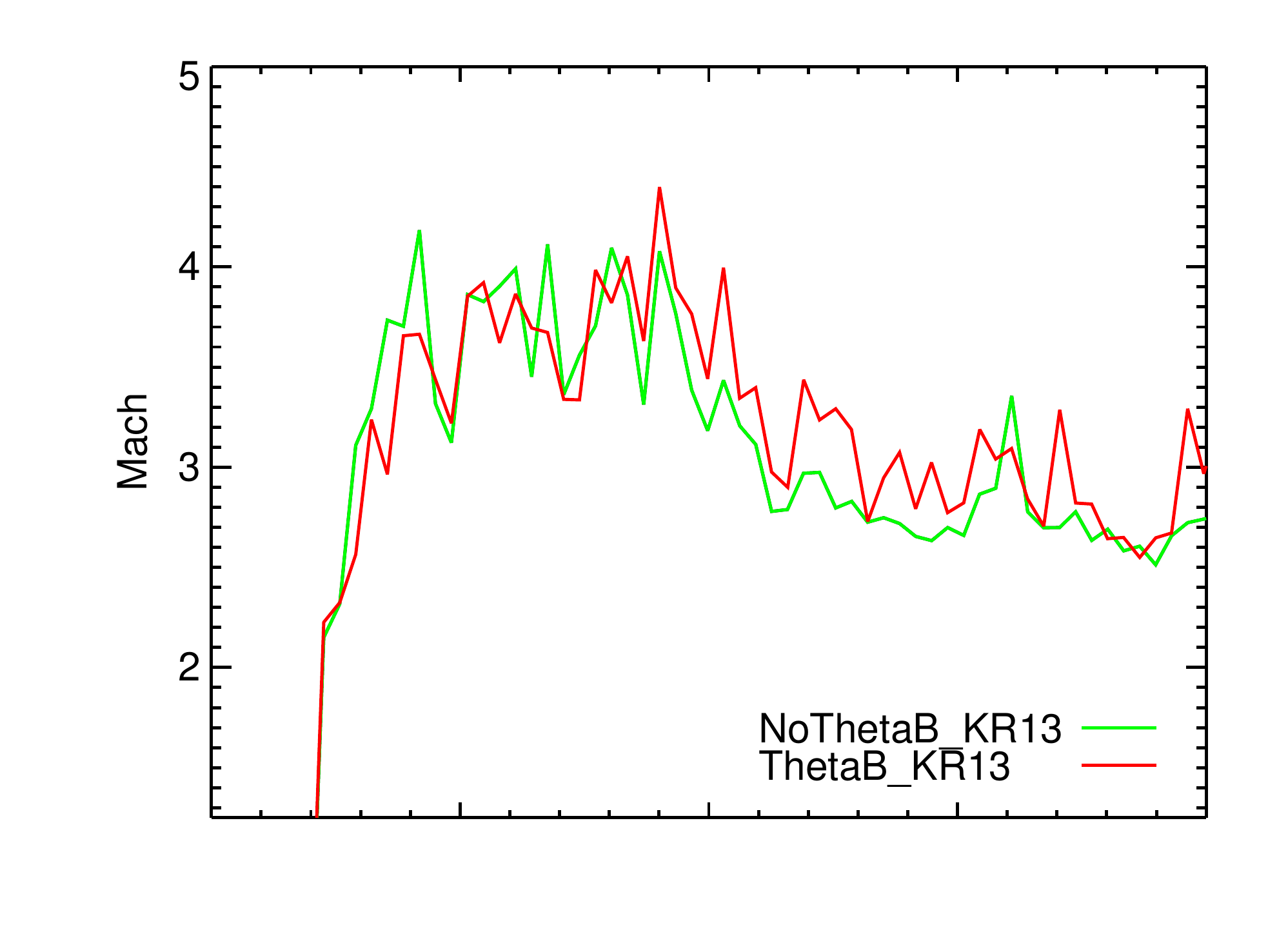}\vspace{-1.25cm}
\centering \includegraphics[width=0.45\textwidth]{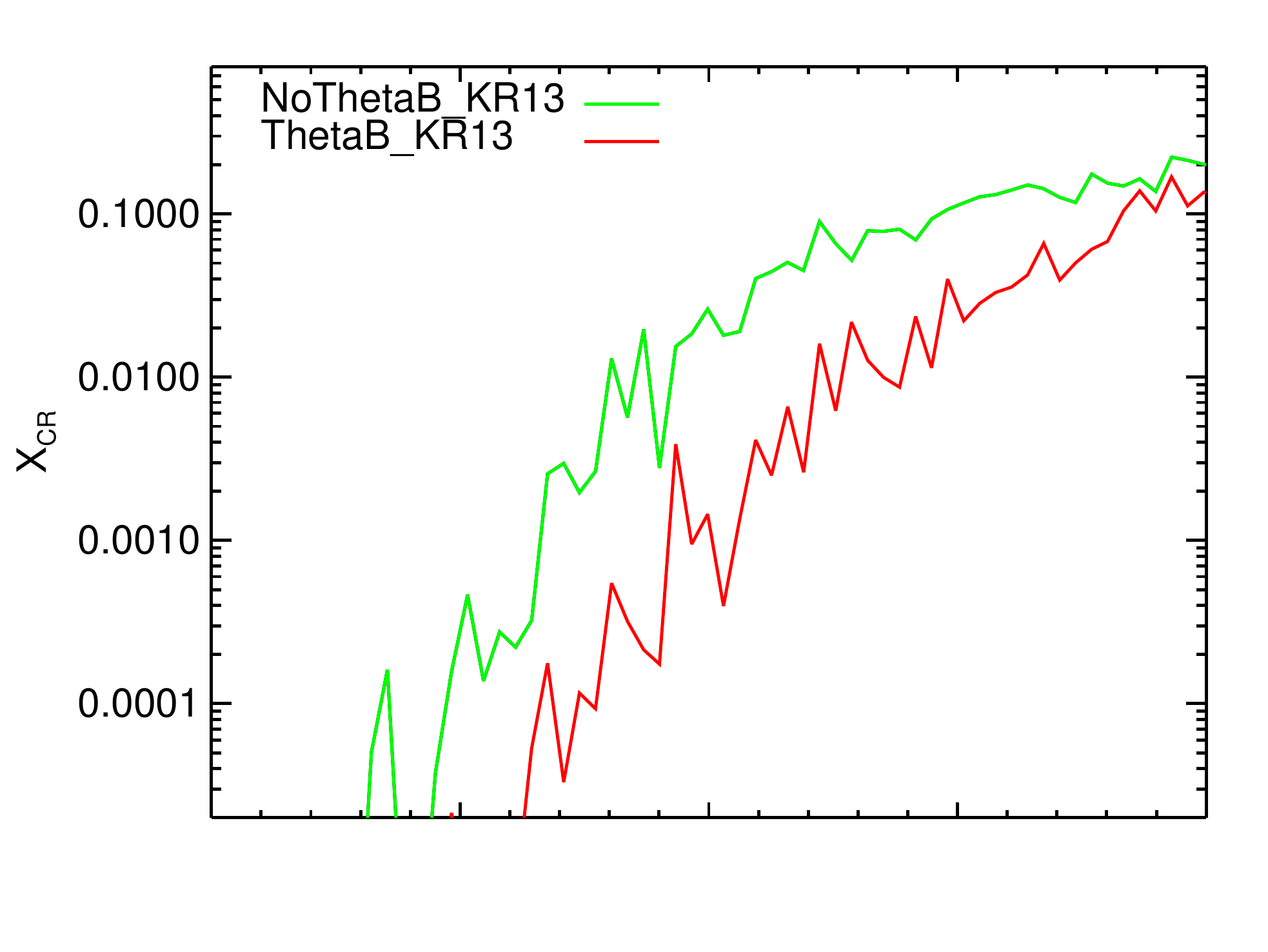}\vspace{-1.25cm}
\centering \includegraphics[width=0.45\textwidth]{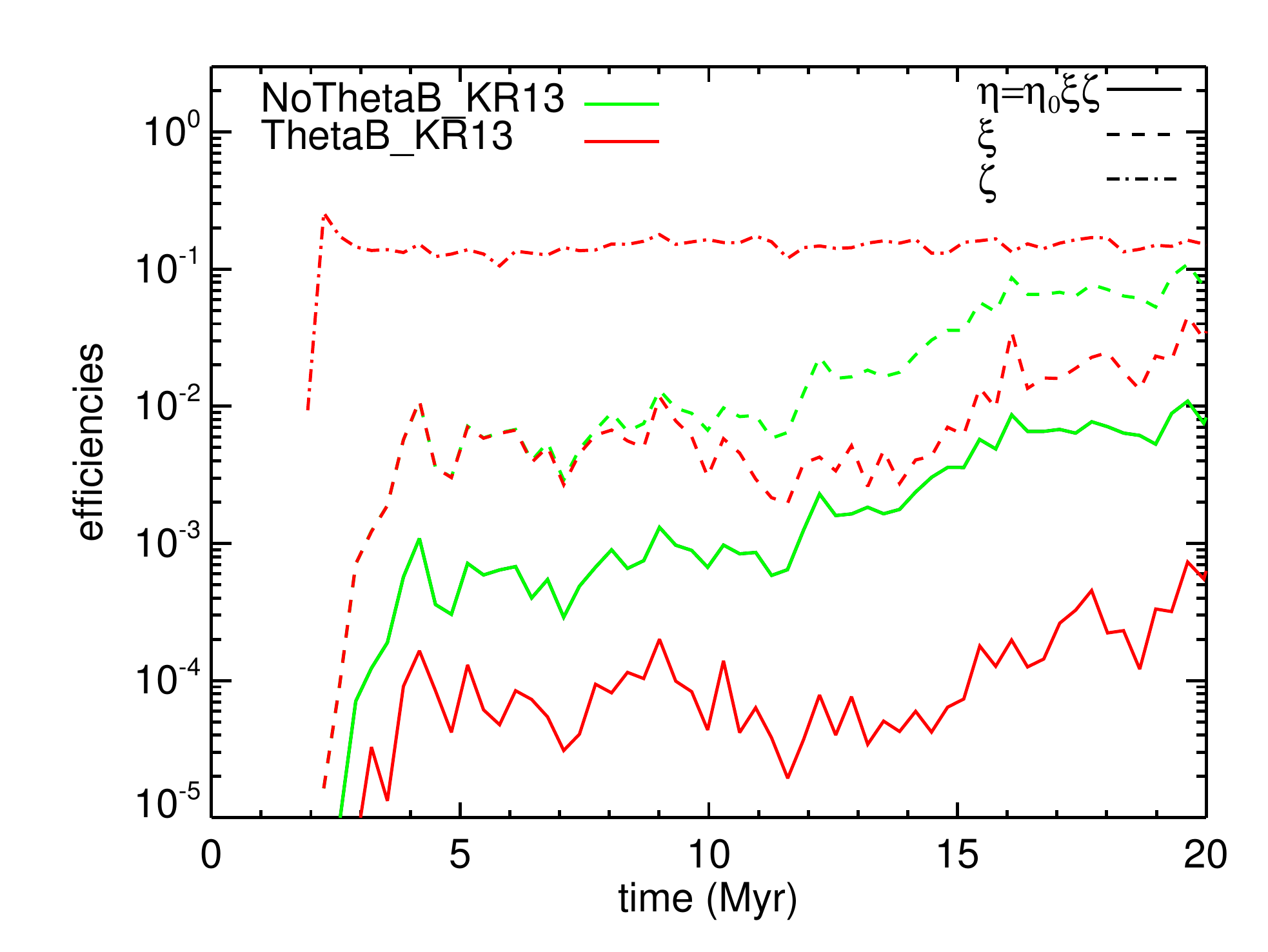}
\caption{Top to bottom: Cosmic-ray flux-weighted mean Mach number $\mathcal{M}$, CR flux-weighted mean CR-to-thermal pressure ratio $X_{\rm CR}$, and dissipated energy flux-weighted efficiencies $\eta=\eta_0\xi(\mathcal{M},X_{\rm CR})\zeta(\theta_{\rm B})$, $\xi(\mathcal{M},X_{\rm CR}),$ and $\zeta(\theta_{\rm B})$ as a function of time for the simulation with (red) and without (green) obliquity dependency. We recall that $\eta_0=0.1$ is used in those simulations and that $\xi$ is a function of $\mathcal{M}$ and $X_{\rm CR}$ as extrapolated from the values of~\cite{kang&ryu13}. }
\label{fig:ismkr13}
\end{figure}

We show here the results of the turbulent box experiments, where this time the efficiency dependency $\xi(\mathcal{M},X_{\rm CR})$ is not assumed to be equal to 1, but varies according to the scaled values of~\cite{kang&ryu13}.
We ran two numerical experiments, free of CR streaming, with and without the magnetic obliquity dependency $\zeta(\theta_{\rm B}),$ called ThetaB\_KR13 and NoThetaB\_KR13, respectively.
We recall that we started with an initial CR pressure of almost  zero   so that $X_{\rm CR}=10^{-10}$ everywhere in the box at time $t=0$, and that a normalisation (maximum) acceleration efficiency $\eta_0=0.1$ was used throughout.

Figure~\ref{fig:ismkr13} shows the evolution of the CR flux-weighted mean value of $\mathcal{M}$ (top panel) and $X_{\rm CR}$ (middle panel), and the evolution of the energy flux-weighted mean acceleration efficiencies (bottom panel) $\eta=\eta_0\xi(\mathcal{M},X_{\rm CR})\zeta(\theta_{\rm B})$, $\xi(\mathcal{M},X_{\rm CR})$, and $\zeta(\theta_{\rm B})$ as a function of time.
The bulk of the CR energy is produced in shocks of $\mathcal{M}\simeq3-4$ with a slight decrease over time.
As CRs are produced, the upstream CR-to-thermal pressure ratio rises to values close to $X_{\rm CR}\simeq 0.1-0.2$ at time $t=20\,\rm Myr$.
The corresponding CR acceleration efficiencies also evolve with time since $\xi$ varies significantly for this range of moderate Mach number as a function of $X_{\rm CR}$ reaching $\xi\simeq 0.03$ and 0.1 at $t=20\, \rm Myr$ for the ThetaB\_KR13 and NoThetaB\_KR13 runs respectively.
In particular, there is an increase between 10 and $20\, \rm Myr$ of the acceleration efficiency by one order of magnitude in both simulations.
The difference between the two simulations is that the obliquity dependent run has a lower overall acceleration efficiency $\eta$ since nearly random magnetic fields (see Fig.~\ref{fig:thetapdf_ism}) reduce the $\zeta$ component to $\simeq 0.2$.
We note that the choice of starting with $X_{\rm CR}=0$ for educative purposes makes these simulations extremely unrepresentative of the ISM of normal galaxies (though it might apply for proto-galaxies),  and delay the build-up of the CR pressure.
Nonetheless, we show that our implementation of the $\mathcal{M}$, $X_{\rm CR}$ (and $\theta_{\rm B}$) dependency  of $\eta$ leads to interesting results in the build-up of the CR pressure through shocks, and might be useful for a broad range of applications.

\section{Conclusion}
\label{section:conclusion}

We have introduced a new modelling of anisotropic CR streaming and dynamical CR shock-acceleration for the AMR code {\sc ramses}~\citep{teyssier02}.
Streaming is solved with a diffusion approach where the diffusion step is performed with a time implicit scheme~\citep{dubois&commercon16}, and can handle complex multi-dimensional problems with non-trivial magnetic field geometries.
CR acceleration at shocks through the DSA mechanism is obtained by accurately detecting shocks, and measuring their Mach number and magnetic obliquity.
We have shown that our numerically CR accelerated solutions faithfully reproduces exact 1D Sod shock tube solutions.
CR-modified 3D Sedov solutions with accelerated CRs have been tested with various background magnetic field configurations (hence, obliquities).
They show very good agreement with previous numerical experiments~\citep{pfrommeretal17} with CRs reducing the effective adiabatic index and slowing down the motion of the shell.
Obliquity dependency of the acceleration leads to a significant modification of the CR distribution in the shell of the Sedov explosion with either a polar or patchy distribution when the coherence length of the background magnetic field is respectively larger or smaller than the bubble size.
This also has consequences on the final shape of the bubble, with a significant elongation of the bubble when the magnetic field has a large field coherence with respect to the bubble size~\citep{paisetal18}.

Finally, the effect of CR streaming and CR  acceleration has been tested in a turbulent box mimicking the motions within the interstellar medium on scales of  tens of pc ~\citep{commerconetal19}.
CRs are produced at shock surfaces and are spread throughout the entire volume by convection and streaming.
CRs have important consequences on the reservoir of cold gas available as they provide a long-term pressure support against compressed material, and streaming substantially modifies the small-scale distribution of CRs, and in turn the clustering of gas.
The obliquity of the field produces a strong suppression of the effective acceleration efficiency, a factor of $\sim2$ beyond the pure random case as a result of the preferential alignment of magnetic fields with shock surfaces.

These new CR physics modules embedded in the {\sc ramses} code make it useful for the  study of the impact of CRs in a wide variety of situations, such as the acceleration of CRs by cosmic shocks, galactic-wide outflows driven by CRs (Dashyan \& Dubois, sub.), the release of CRs in galaxy clusters by active galactic nuclei, studies of supernova remnants, and the release of CRs in the supernova-driven turbulence of the ISM, which we defer to future work.

\begin{acknowledgements}

We thank G. Dashyan, M. Lemoine, C. Pfrommer, R. Teyssier, and A. Wagner for enlightening discussions.
We warmly thank S. Rouberol for smoothly running the Horizon cluster on which several simulations were run.
This work was supported by the the ANR grant LYRICS (ANR-16-CE31-0011) and the CNRS programs `Programme National de Cosmologie et Galaxies' (PNCG) and `Physique et Chimie du Milieu Interstellaire' (PCMI).
\end{acknowledgements}

\bibliographystyle{aa}
\bibliography{author}

\appendix

\section{Effect of perpendicular diffusion on streaming}
\label{appendix:sinusoid2d_kiso}

\begin{figure}
\centering \includegraphics[width=0.24\textwidth]{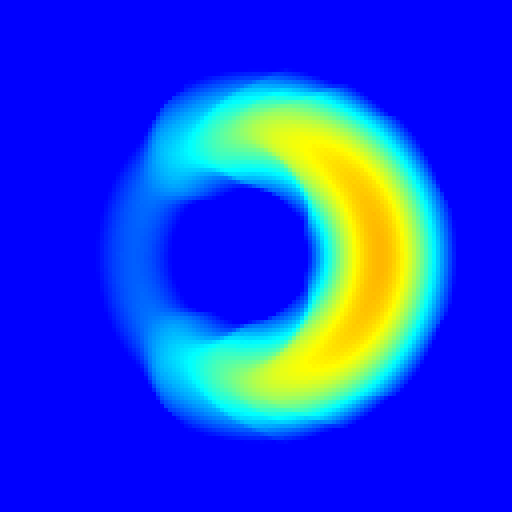}
\centering \includegraphics[width=0.24\textwidth]{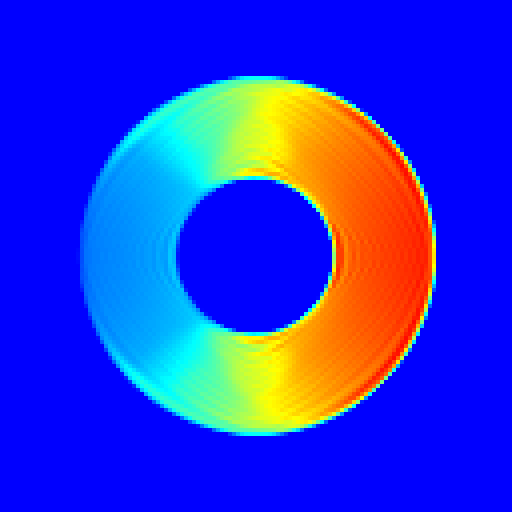}
\caption{Energy density maps at $t=0.02$ for the same set-up as for the 2D sinusoid loop described in Section~\ref{section:sinusoid2d} with $f_{\rm iso}=10^{-1}$ (left) or $f_{\rm iso}=10^{-3}$ (right).}
\label{fig:loop_sinusoid_comp}
\end{figure}

Here we vary the value of the isotropic component of the streaming diffusion term from $f_{\rm iso}=10^{-3}$ to $f_{\rm iso}=10^{-1}$ (to be compared with the value of $f_{\rm iso}=10^{-2}$ used by default in section~\ref{section:sinusoid2d}) with respect to pure anisotropy.
Figure~\ref{fig:loop_sinusoid_comp} shows that increasing the value of $f_{\rm iso}$ to $10^{-1}$  leads to more diffusion outside of the loop, which decreases the values of the maximum, while $f_{\rm iso}=10^{-3}$ produces numerically driven finger-like features but allows  a more contained CR distribution in the loop.

\section{Tabulated values of $\xi(\mathcal{M},X_{\rm CR})$}
\label{app:accKR13}

\begin{table*}[h]
 \caption[]{\label{table:KR13}Acceleration efficiencies interpolated values of $\xi(\mathcal{M},X_{\rm CR})$ from~\cite{kang&ryu13}. }
\centering\begin{tabular}{l|ccccccc}
 \hline \hline
   &
  $X_{\rm CR}=0$ &
  $X_{\rm CR}=0.025$ &
  $X_{\rm CR}=0.05$ &
  $X_{\rm CR}=0.1$ &
  $X_{\rm CR}=0.2$ &
  $X_{\rm CR}=0.5$ &
  $X_{\rm CR}=1$
 \\ \hline
$\mathcal{M}=2$    & $4.44\times10^{-4}$ & $3.80\times 10^{-2}$ & $7.55\times10^{-2}$ & $1.51\times10^{-1}$ & $3.01\times10^{-1}$ & $7.51\times10^{-1}$ & $1.00$                      \\
$\mathcal{M}=3$    & $2.66\times10^{-2}$ & $1.47\times 10^{-1}$ & $2.66\times10^{-1}$ & $5.06\times10^{-1}$ & $9.86\times10^{-1}$ & $1.00$                       & $1.00$                      \\
$\mathcal{M}=4$    & $2.00\times10^{-1}$ & $4.11\times 10^{-1}$ & $6.22\times10^{-1}$ & $9.08\times10^{-1}$ & $1.00$                      & $1.00$                       & $1.00$                      \\
$\mathcal{M}=5$    & $4.44\times10^{-1}$ & $5.60\times 10^{-1}$ & $6.76\times10^{-1}$ & $9.08\times10^{-1}$ & $1.00$                      & $1.00$                       & $1.00$                      \\
$\mathcal{M}=7$    & $6.66\times10^{-1}$ & $7.33\times 10^{-1}$ & $8.00\times10^{-1}$ & $9.33\times10^{-1}$ & $1.00$                      & $1.00$                       & $1.00$                      \\
$\mathcal{M}=10$  & $8.66\times10^{-1}$ & $8.89\times 10^{-1}$ & $9.10\times10^{-1}$ & $9.55\times10^{-1}$ & $1.00$                      & $1.00$                       & $1.00$                      \\
$\mathcal{M}=20$  & $1.00$                      & $1.00$                       & $1.00$                      & $1.00$                      & $1.00$                      & $1.00$                      & $1.00$                \\
$\mathcal{M}=30$  & $1.00$                      & $1.00$                       & $1.00$                      & $1.00$                      & $1.00$                      & $1.00$                      & $1.00$                \\
$\mathcal{M}=50$  & $1.00$                      & $1.00$                       & $1.00$                      & $1.00$                      & $1.00$                      & $1.00$                      & $1.00$                \\
$\mathcal{M}=100$  & $1.00$                      & $1.00$                       & $1.00$                      & $1.00$                      & $1.00$                      & $1.00$                      & $1.00$                \\
\hline
\end{tabular}
\end{table*}

Table~\ref{table:KR13} shows the tabulated values of~\cite{kang&ryu13} renormalised to 1 (see section~\ref{section:cracc} for details).

\section{Shock numerical broadening}
\label{appendix:shock_broadening}

\begin{figure}
\centering \includegraphics[width=0.4\textwidth]{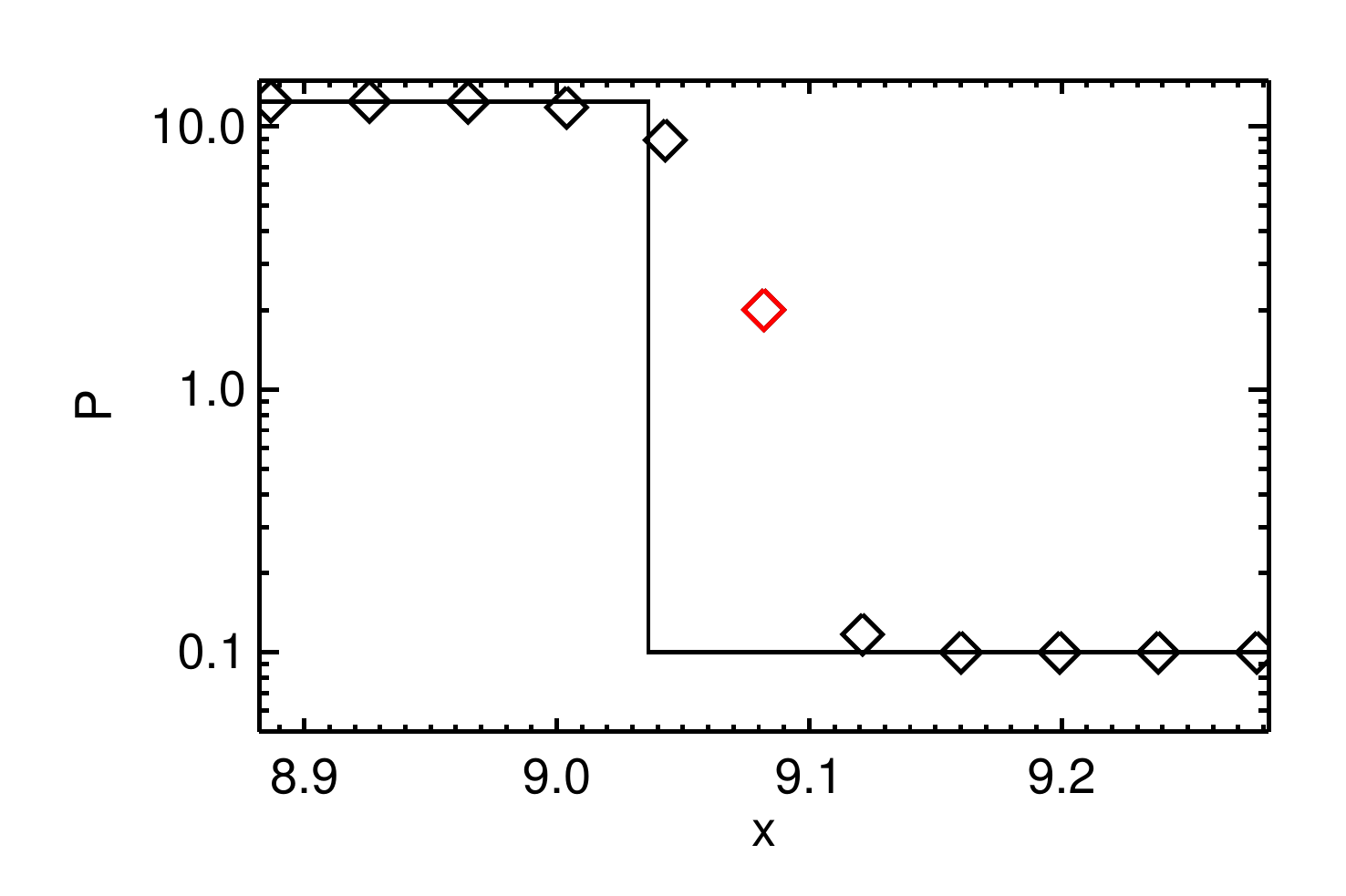}\vspace{-.5cm}
\centering \includegraphics[width=0.4\textwidth]{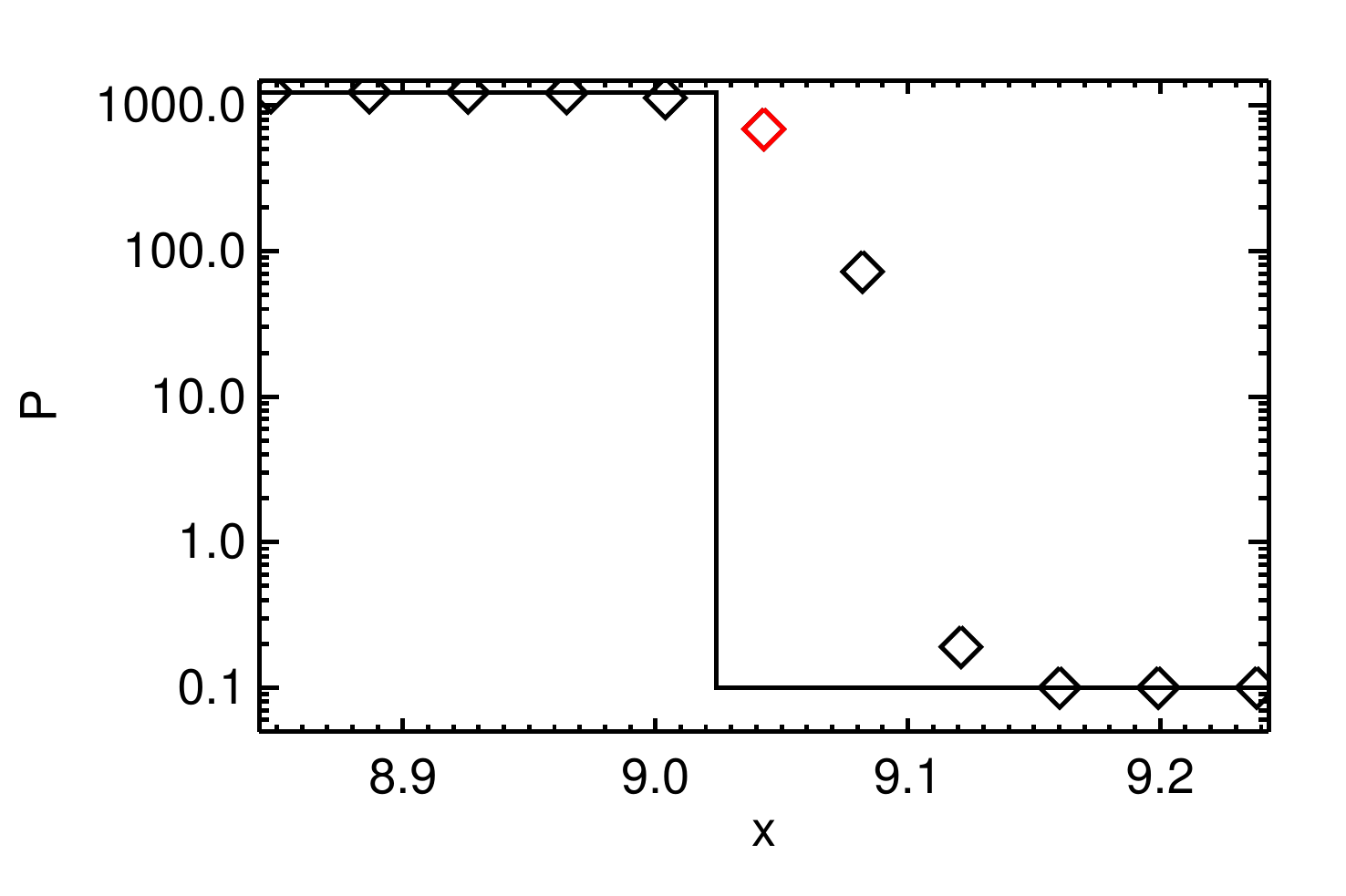}\vspace{-.5cm}
\centering \includegraphics[width=0.4\textwidth]{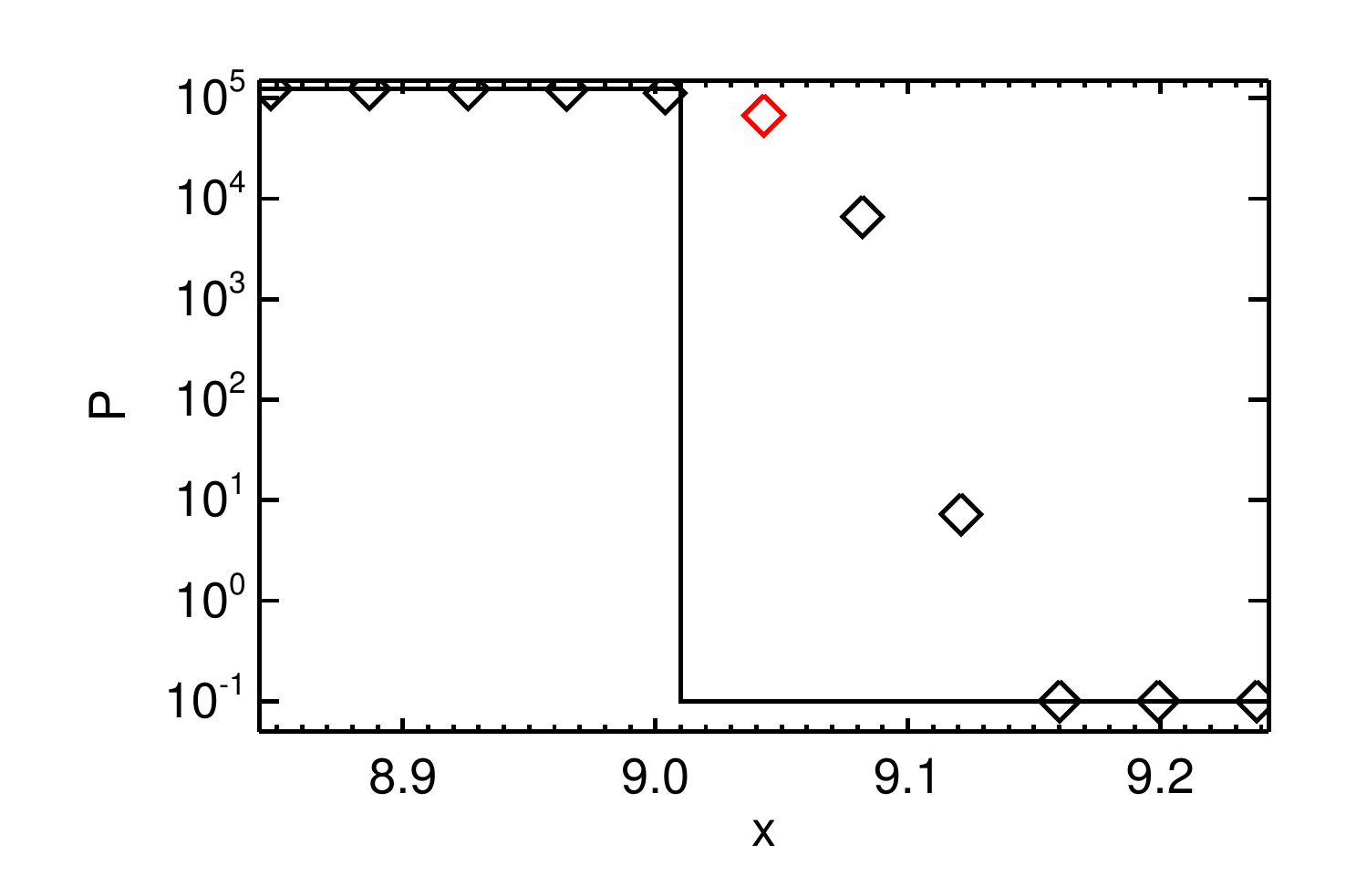}
\caption{Pressure profiles at time $t\simeq3.5/\mathcal{M}$ around the shock discontinuity for the Mach $\mathcal{M}=$10 (top), 100 (middle), and 1000 (bottom) experiments. The result of the numerical solution is shown as diamonds;  the red symbol highlights the position of the shock cell given by a shock finder algorithm. The solid line is the exact numerical solution. We see that the numerical shock tends to broaden with increasing Mach number, and given the largest error made on the post- and pre-shock regions, the error on the evaluated Mach number becomes larger for a small kernel ($n_{\rm cell,max}=2$).}
\label{fig:shock_broadening}
\end{figure}

We show in Fig.~\ref{fig:shock_broadening} a zoomed-in view of the shock discontinuity for the Sod shock tube experiments described in Section~\ref{sec:sodnocr} (i.e. without CRs) and for the three different Mach numbers $\mathcal M=10$, 100, and 1000.
Instead of a pure discontinuity (the exact solution is shown as a solid line) the numerical shock  is broadened by numerical diffusion with typically 4-5 cells;  the number of cells in the discontinuity to match the exact pre- and post-shock pressures increases with the value of the Mach number, and given the quadratic increase in pressure jump with Mach number, any error is strongly amplified.
In the strongest shock example shown in the bottom panel, using only two cells away from the shock would lead to underestimating the Mach number by a factor of 10 (Mach number scales with $R_{\rm P}^{1/2}$ and the upstream value two cells away from the shock is $\simeq100$ times that of the true value).

\section{Random magnetic fields}
\label{app:brandom}

In order to set a random magnetic field fulfilling the $\nabla . \vec B=0$ constraint, we first set up a random potential vector on the nodes of a cartesian grid of arbitrary resolution $n_{\rm pot}^3$ cells (there are actually $(n_{\rm pot}+1)^3$ values of potential vectors drawn at nodes of the $n_{\rm pot}^3$ sampling cells), with the right-, top-, and back-most boundaries being replicates of the left-, bottom-, and front-most boundaries to ensure the correct periodicity of the (staggered) magnetic field.
In the cases simulated in this paper the AMR cell size is smaller than or equal to $1/n_{\rm pot}$, which means that the vector potential is the trilinear interpolation of the surrounding node vector potentials projected along the AMR cell edge.
Once these reconstructed vector potentials are obtained along AMR cell edges, the staggered magnetic field (one B-field perpendicular to each face of AMR cells) is obtained by taking the rotational of the potential vector of the face-surrounding edges.
This procedure guarantees that the magnetic field is random, $\nabla.\vec B=0$, and the consistency of the coarse-to-fine values of the B-field. 
We note that we took the initial random potential vector as a white noise vector, but this can be modified to account for any given spectrum of the vector potential (or magnetic field), and to obtain any desired shape of the magnetic power spectrum, as the power spectrum of $\vec{B}$ scales as $k$ (i.e. the wave number) times the power spectrum of $\vec{A}$.

\end{document}